\documentclass[article]{jss}

\input{header.sty}

\usepackage{thumbpdf,lmodern}

\usepackage{multirow, booktabs} 
\usepackage{tikz} 
\usetikzlibrary{shapes.geometric}
\usetikzlibrary{positioning,fit,calc}

\usepackage{framed}

\newcommand{\class}[1]{`\code{#1}'}
\newcommand{\fct}[1]{\code{#1()}}

\author{Jakob A. Dambon\\Lucerne University of \\Applied Sciences and Arts
   \And Fabio Sigrist\\Lucerne University of \\Applied Sciences and Arts
   \And Reinhard Furrer\\University of Zurich}
\Plainauthor{Jakob A. Dambon, Fabio Sigrist, Reinhard Furrer}

\title{\pkg{varycoef}: An \proglang{R} Package for Gaussian Process-based Spatially Varying Coefficient Models}
\Plaintitle{varycoef: An R Package for Gaussian Process-based Spatially Varying Coefficient Models}
\Shorttitle{\pkg{varycoef}: GP-based SVC Models}

\Abstract{
  Gaussian processes (GPs) are well-known tools for modeling dependent data with applications in spatial statistics, time series analysis, or econometrics. In this article, we present the \proglang{R} package \pkg{varycoef} that implements  estimation, prediction, and variable selection of linear models with spatially varying coefficients (SVC) defined by GPs, so called GP-based SVC models. Such models offer a high degree of flexibility while being relatively easy to interpret. Using \pkg{varycoef}, we show versatile applications of (spatially) varying coefficient models on spatial and time series data. This includes model and coefficient estimation with predictions and variable selection. The package uses state-of-the-art computational statistics techniques like parallelization, model-based optimization, and covariance tapering. This allows the user to work with (S)VC models in a computationally efficient manner, i.e., model estimation on large data sets is possible in a feasible amount of time.}

\Plainkeywords{covariance tapering, dependent data, model-based optimization, spatial statistics, (penalized) maximum likelihood estimation, variable selection}

\Address{
  Jakob A. Dambon\\
  Department of Mathematics\\
  Faculty of Science\\
  University of Zurich\\
  Winterthurerstr.~190\\
  8057 Zurich, Switzerland\\
  E-mail: \email{jakob.dambon@math.uzh.ch}\\
  \emph{and}\\
  Institute of Financial Services Zug\\
  Lucerne School of Business\\
  Lucerne University of Applied Sciences and Arts\\
  Suurstoffi~1\\
  6343 Rotkreuz, Switzerland\\
  \\
  Fabio Sigrist\\
  Institute of Financial Services Zug\\
  Lucerne School of Business\\
  Lucerne University of Applied Sciences and Arts\\
  Suurstoffi~1\\
  6343 Rotkreuz, Switzerland\\
  \\
  Reinhard Furrer\\
  Department of Mathematics\\
  \emph{and}\\
  Department of Computational Sciences\\
  Faculty of Science\\
  University of Zurich\\
  Winterthurerstr.~190\\
  8057 Zurich, Switzerland\\
}

\begin{document}




\section{Introduction} \label{sec:intro}

Spatially varying coefficients (SVC) provide a flexible and interpretable approach to extend linear models to spatial data. There are various methodologies on how to estimate and make predictions for SVC models. To name a few, geographically weighted regression (GWR) introduced by \citet{Brundson1998} or Bayesian SVC processes by \citet{Gelfand2003} are popular examples. Today, there are several \proglang{R} packages and other software implementations for SVC modeling available; each with individual focuses in their respective framework. A thorough comparison between all of them is beyond the scope of this work. Instead, we provide a rough outline of existing software implementations. 

In \proglang{R} \citep{R}, GWR has been implemented in packages like \pkg{GWmodel} \citep{R:GWmodel}, \pkg{spgwr} \citep{R:spgwr}, and \pkg{gwrr} \citep{R:gwrr}. A detailed comparison between those packages can be found in \citet{R:GWmodel}. Bayesian SVC modeling is implemented in the \proglang{R} packages \pkg{spTDyn} \citep{Bakar2016} and \pkg{spBayes} \citep{Finley2015, Finley2020}. Both packages use Markov chain Monte Carlo (MCMC) sampling algorithms and are rather restricted in the number of observation locations. Another Bayesian method to estimate SVC models uses the explicit stochastic partial differential equation (SPDE, \citealp{Lindgren2011}) link between Gaussian fields and Gaussian Markov random fields (GMRF, \citealp{Rue2005}). Using integrated nested Laplace approximation (INLA) implemented in the \proglang{R} package \pkg{INLA} \citep{R:INLA}, one can estimate SVC models for data sets with a large number of observations. However, the number of hyper parameters and therefore varying coefficients is limited \citep{INLA:review2016}. Finally, spatially varying coefficients can be modeled using splines. Available options include the packages \pkg{mgcv} \citep{Wood2017} and \pkg{mboost} \citep{R:mboost}. In other programming languages some of the above mentioned methodologies are available, too. For instance, the \proglang{Python} spatial analysis library \pkg{PySAL} \citep{Rey2010} implements Bayesian SVC processes by \citet{Gelfand2003}. GWR is also available in \pkg{PySAL} and geographic information system (GIS) software like \proglang{ArcGIS} or \proglang{GRASS}.

This article discusses SVC models where each coefficient is defined by a Gaussian process (GP, \citealp{Rasmussen2005}), so called GP-based SVC models. The proposed model is similar to Bayesian SVC processes by \citet{Gelfand2003}, but with some specific assumptions on the model, we can provide a computationally efficient way of estimating GP-based SVC models using a classical maximum likelihood estimation (MLE) approach \citep{JAD2020}. Therefore, in contrast to all of the methodologies and software implementations above, the \proglang{R} package \pkg{varycoef} implements a frequentist approach for SVC modeling using Gaussian processes. Additionally, while all of the above SVC modeling implementations from above are either limited in the number of observations or spatially varying coefficients, \pkg{varycoef} has been developed for to work well with large data sets or a moderate number of varying coefficients. Over time, the methodology as well as the corresponding \proglang{R} package \pkg{varycoef} have been augmented continuously. For instance, the package now implements a joint variable selection procedure for GP-based SVC models \citep{JAD2021} using penalized maximum likelihood estimation (PMLE). Another new feature of \pkg{varycoef} is the support of different types of covariance functions and the idea of SVC models has been generalized to work different types of dependent data such as time series. The goal of this article is to present the current state of the package \pkg{varycoef} with its versatile applications.

The rest of this article is structured as follows. In Section~\ref{sec:models} we introduce GP-based SVC models in their original form to be used by \pkg{varycoef}. Section~\ref{sec:MLE} covers the MLE of GP-based SVC models including prediction methods. The variable selection using PMLE is discussed and showcased in Section~\ref{sec:PMLE}. In particular, we move from a classical application using spatial data and show an application on time series data. Section~\ref{sec:summary} summarizes this work.



\section{GP-based SVC Models} \label{sec:models}

We commence with a formal introduction of Gaussian processes before extending the linear regression models to GP-based SVC models. 

\subsection{Gaussian Processes}

Gaussian processes are widely used for modeling dependency structures. Applications can be found in -- but are not limited to -- spatial statistics \citep{Gelfand2016, Banerjee2008, Datta2016}, econometrics \citep{Wu2014}, and time series modeling \citep{Roberts2013}. Similarly to a normal distribution, a GP is defined as an infinite-dimensional process with a mean function $\bmu$ and a covariance function $c$,
\begin{align*}
  \bmu (\cdot)&: D \rightarrow \mathbb R, \\
  c(\cdot, \cdot; \btheta)&: D \times D \rightarrow \left[0, \infty \right),
\end{align*}
for some domain $D \subset \mathbb{R}^d, d\geq 1$, and covariance parameters $\btheta$. In this work, we restrict ourselves to constant mean functions and isotropic covariance functions. That is, the covariance function is only depending on the distance of its arguments $u = \| \s - \s' \|$, where $\|\cdot \|$ denotes the Euclidean distance and $\s, \s' \in D$.

\begin{figure}[t!]
\centering
\includegraphics{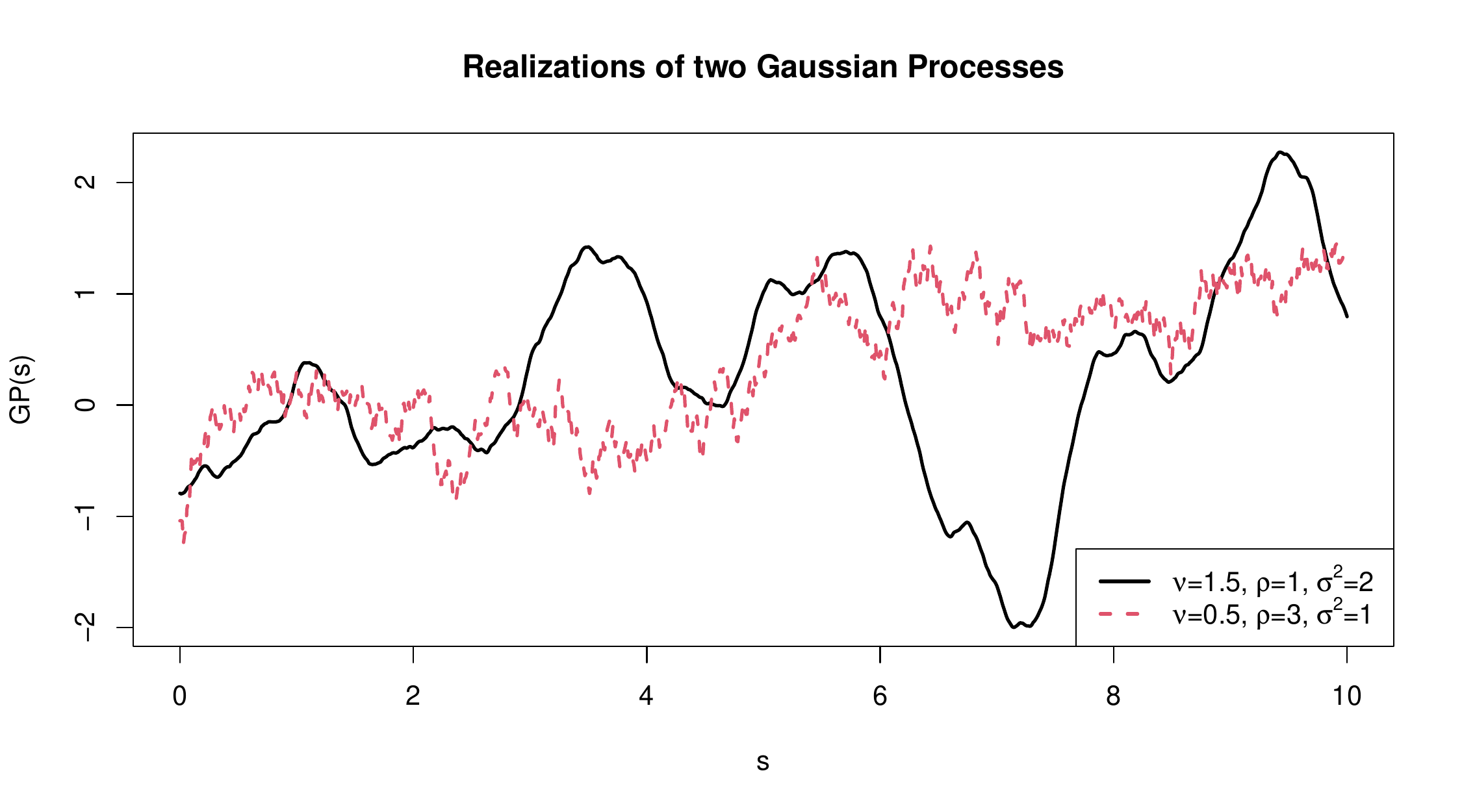}
\caption{\label{fig:twoGPs} Two zero-mean Gaussian processes with Mat\'ern covariance functions and distinct parametrizations. The observation location $\s_i$ are defined equidistantly on $[0, 10] \subset \IR$ with distance 0.01, i.e., $n = 1001$. One can clearly see the effects of the parameters $\nu, \rho$, and $\sigma^2$ on the smoothness, the range of dependence, and the variance of the curves, respectively.}
\end{figure}

Popular examples of covariance functions are given by the Mat\'ern or generalized Wendland covariance class. In the isotropic case, former one is defined as
\begin{align}
  c: \left[0, \infty \right) &\to \left[0, \infty \right), \nonumber \\
   u &\mapsto \sigma^2 \frac{2^{1- \nu} }{\Gamma(\nu)} \left(\sqrt{2\nu} \frac{u}{\rho} \right)^\nu K_\nu \left(\sqrt{2\nu} \frac{u}{\rho} \right) \label{eq:matern},
\end{align}
where $\sigma^2 \geq 0$ is the variance, $\rho > 0$ is the range, $\nu > 0$ is the smoothness, and $K_\nu$ is the modified Bessel function of second kind and order $\nu$. Setting the smoothness $\nu$ to some specific values simplifies the bulky formula of \eqref{eq:matern} to simple functions like the exponential $c(u) = \sigma^2 \exp (-u/\rho)$ for $\nu = 1/2$ or the squared exponential $c(u) = \sigma^2 \exp (-u^2/\rho)$ for $\nu = \infty$. We let the definition of the covariance function up to the user and only assume that the covariance parameters $\btheta := (\rho, \sigma^2)$ are unknown. Both are essential to interpret the estimated Gaussian process as the range $\rho$ provides a measure of spatial dependence and the variance $\sigma^2$ gives the volatility of the Gaussian process. We provide examples of Gaussian processes defined by zero-means and different covariance functions of Mat\'ern class. They are given on $[0, 10] \subset \IR$, i.e., in $d = 1$ dimension. Throughout this article as well as in \pkg{varycoef}, we use the package \pkg{RandomFields} \citep{R:RandomFields} to sample Gaussian processes. The two sampled processes are depicted in Figure~\ref{fig:twoGPs}.

\subsection{Spatially Varying Coefficient Models}

Let $n$ be the number of observations and let $p$ be the number covariates given by $\x^{(j)} := (x_1^{(j)}, ..., x_n^{(j)}) \in \mathbb R^n, j = 1,..., p$. With the responses $\y = (y_1, ..., y_n) \in \mathbb R^n$, errors $\bvarepsilon := (\varepsilon_1, ..., \varepsilon_n) \sim \cN_n (\0, \tau^2\I_n)$ where $\tau>0$, and coefficients $\bbeta = (\beta_1, ..., \beta_p) \in \mathbb R^p$ the linear model is given by:
\begin{align*}
 y_i = \beta_1 x_i^{(1)} + ... + \beta_p x_i^{(p)} + \varepsilon_i.
\end{align*}
Spatially varying coefficient models generalize the linear model. In a classical context of regressing spatial data sets, one considers the observation locations $\s_i \in D$ that are associated to each observed sample $i$. Here, one usually assumes $d = 2$ and that the observation locations $\s_i$ do not necessarily have to be distinct. A general SVC model is then given by:
\begin{align*}
 y_i = \beta_1(\s_i) x_i^{(1)} + ... + \beta_p(\s_i) x_i^{(p)} + \varepsilon_i.
\end{align*}
It is a this point where the above mentioned methodologies to estimate the SVC model differ depending on the assumption of the coefficients $\bbeta_j(\cdot)$. In our case, we assume fixed effects $\mu_j$ and random effects $\bfeta_j(\cdot)$ defined by zero-mean Gaussian processes with an isotropic covariance function $c_j(\cdot; \btheta_j)$ to model the spatial structures of the coefficients, i.e, we have $\bbeta_j(\cdot) \approx \mu_j + \bfeta_j(\cdot)$. Additionally, we assume prior mutual independence between all $\bfeta_j(\cdot)$. 

For a finite set of observations $\s = (\s_1, ..., \s_n)$ the Gaussian processes $\bfeta_j(\cdot)$ from above reduce to zero-mean normal distributions. Therefore, we can write the GP-based SVC model as a linear mixed model: 
\begin{align}\label{eq:GPSVC}
  \y = \X \bmu + \W \bfeta(\s) + \bvarepsilon.
\end{align}
The full derivation of \eqref{eq:GPSVC} is given in \citet{JAD2020}. In the resulting model, the first term on the right hand side is the data matrix $\X = \bigl(\x^{(1)} | ... | \x^{(p)}\bigr) \in \mathbb R^{n\times p}$ associated with the fixed effects $\bmu = (\mu_1, ..., \mu_p)$. The random effects and its corresponding covariates are given in the second term. Not every fixed effect covariate $\x^{(j)}$ has to be associated with a random effect or vice versa. Therefore, we denote the $q$ random effect covariates by $\w^{(k)} \in \IR^n$ for $k = 1,..., q$ and the data matrix $\W = \bigl(\diag \w^{(1)} | ... |\diag \w^{(q)} \bigr) \in \IR^{q \times nq}$. The random effects are contained in $\bfeta(\s) \in \IR^{nq}$ which is the sole component modeling the spatially varying relationship depending on the locations $\s$. Individual zero-mean spatially varying coefficients are defined as $\bfeta_k(\cdot) \sim \cG\cP \bigl(0, c_k(\cdot; \btheta_k)\bigr)$. For observation locations $\s$, they reduce to a normal distribution with $\bfeta_k(\s) \sim \cN_n(\0, \bSigma_k)$, where $\bigl(\bSigma_k\bigr)_{lm} := c_k(\|\s_l - \s_m\|; \btheta_k)$. The random effect $\bfeta(\s)$ is the joint effect over all individual Gaussian processes, i.e., $\bfeta(\s) = \bigl(\bfeta_1(\s), ..., \bfeta_q(\s)\bigr) \sim \cN_{nq} (\0, \bSigma)$ with joint block covariance matrix $\bSigma = \diag (\bSigma_1, ..., \bSigma_q)$. Finally, we add the errors $\bvarepsilon$, also called the nugget in spatial modeling. 

\begin{figure}
\centering
\includegraphics{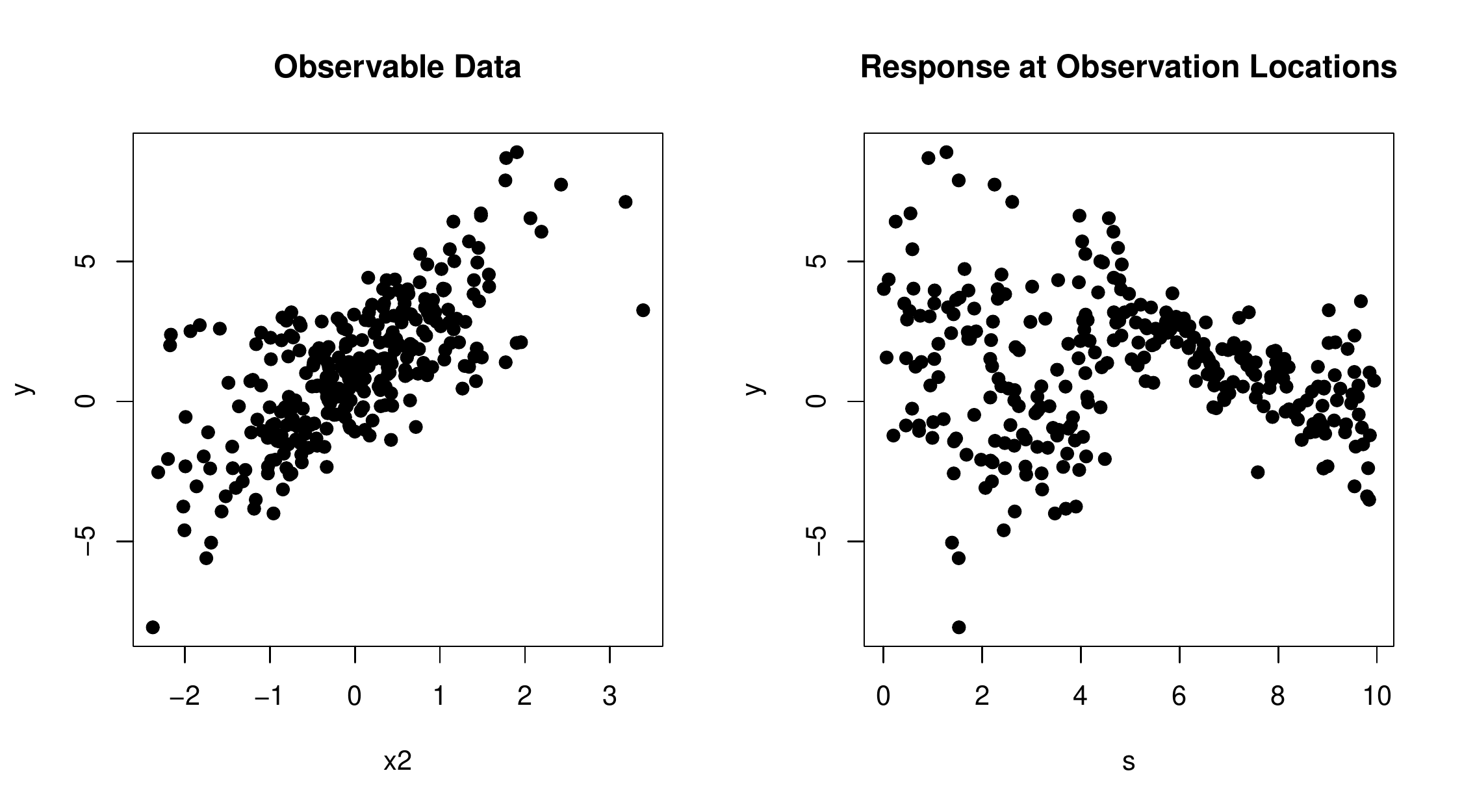}
\caption{\label{fig:simpleSVCmodel} Sampled data using function \fct{sample\_fullSVC} with spatially varying coefficients for the intercept $\x^{(1)}$ and covariate $\x^{(2)}$, respectively. Here, we only display the data without the true coefficients as latter ones are unknown in real world applications, too.}
\end{figure}

Inspired by the example in Figure~\ref{fig:twoGPs}, we sample data under the assumption of an GP-based SVC model. The package \pkg{varycoef} provides the function \fct{sample\_fullSVC} to sample data from a GP-based SVC model with observations on the real line, i.e., $d = 1$. In total we consider $n = 300$ observations with i.i.d.\ locations $s_i \sim \cU\bigl( [0, 10] \bigr)$. It is called a full SVC model since each covariate is associated with a spatially varying coefficient, i.e., $p = q$ and $\x^{(j)} = \w^{(j)}$. Here, \fct{sample\_fullSVC} gives an intercept $\x^{(1)} = \1_n$ and $\x^{(2)} \sim \cN_n(\0_n, \I_{n\times n})$. The mean and covariance parameters are provided in the code below. The resulting data, i.e., the response $\y$ and the covariate $\x^{(2)}$ are depicted in Figure~\ref{fig:simpleSVCmodel}.
\begin{Schunk}
\begin{Sinput}
R> library(varycoef)
R> set.seed(123)
R> # SVC parameters
R> df.pars <- data.frame(var = c(2, 1), scale = c(0.5, 1), mean = c(1, 2))
R> # nugget standard deviation
R> tau <- 0.5
R> # sample locations
R> n <- 300
R> s <- sort(runif(n, min = 0, max = 10))
R> # sample SVCs and data
R> SVCdata <- sample_fullSVC(
+    df.pars = df.pars, nugget.sd = tau, locs = s, cov.name = "mat32")
\end{Sinput}
\end{Schunk}

\subsection{Optimization of Likelihood}

Our goal is to estimate the parameters of interest $\bomega := (\btheta, \bmu)$ with all covariance parameters given by $\btheta = (\rho_1, \sigma_1^2, ..., \rho_q, \sigma^2_q, \tau^2)$. We rely on maximum likelihood estimation, where the multivariate normal distribution and corresponding log-likelihood of the GP-based SVC model \eqref{eq:GPSVC} is given by:
\begin{align}
  \Y &\sim \cN_n\left( \X \bmu, \bSigma_\Y(\btheta)  := \sum_{k = 1}^q \left(\w^{(k)} {\w^{(k)}}^{\top} \right) \odot \bSigma_k + \tau^2 \I_{n\times n}\right), \label{eq:normalY}\\
  \ell(\bomega) &= -\frac{1}{2} \left( n\log (2\pi) + 
		\log \det \bSigma_{\Y} (\btheta) + \left(\y - \X \bmu \right)^\top \bSigma_{\Y}(\btheta)^{-1}\left(\y - \X \bmu \right) \right). \label{eq:ll}
\end{align}

\section{Implementation of MLE}\label{sec:MLE}

\subsection{Control Parameters}

Due to their high modularity, there are several components to define a GP-based SVC model as well as the respective MLE. Here, the function \fct{SVC\_mle\_control} sets several control parameters, which we go through step by step. 

\subsubsection{Covariance Function}

The covariance functions $c_k$ used to define $\bSigma_k$ play a major role in the definition of the likelihood, i.e., the objective function. The argument \code{cov.name} takes a string to define the covariance function $c_k$. The list of supported covariance functions is given in Table~\ref{tab:covfun}. Note that we assume the same covariance function for each Gaussian process and that the covariance function can be written as $c(u; \btheta) = \sigma^2 r(u/\rho)$, i.e., it is given by a correlation function $r(h)$ and only has the range $\rho$ and variance (also called partial sill) $\sigma^2$ as parameters.
\begin{Schunk}
\begin{Sinput}
R> # setting covariance function to Matern with smoothness nu = 3/2
R> SVC_mle_control(cov.name = "mat32")
\end{Sinput}
\end{Schunk}

\begin{table}[t!]
\centering
\renewcommand{\arraystretch}{1.4}
\begin{tabular}{lllll}
\hline
\code{cov.name} & \textbf{Name} & \textbf{Family} & \textbf{Compact} & \textbf{Correlation function} $r(h)$\\ \hline
\code{"exp"}    & Exponential & Mat\'ern ($\nu = 1/2$)  & \code{FALSE} & $\exp(-h)$ \\
\code{"mat32"}  &  \multicolumn{2}{c}{Mat\'ern ($\nu = 3/2$)}  & \code{FALSE} & $\bigl( 1 + \sqrt{3}h\bigr)\exp(-\sqrt{3}h)$ \\
\code{"mat52"}  &  \multicolumn{2}{c}{Mat\'ern ($\nu = 5/2$)}  & \code{FALSE} & $\bigl( 1 + \sqrt{5}h + 5h^2/3\bigr)\exp(-\sqrt{5}h)$ \\
\code{"sph"}    & Spherical   & --          & \code{TRUE} & $\bigl[ 1- 3h/2  + h^3/2 \bigr]_{+}$ \\
\code{"wend1"}    & \multicolumn{2}{c}{Wendland ($\kappa = 1$)} & \code{TRUE} & $[1-h]_{+}^4 (4h+1)$ \\
\code{"wend2"}    & \multicolumn{2}{c}{Wendland ($\kappa = 2$)} & \code{TRUE} & $[1-h]_{+}^6 (35h^2/3 + 6h+1)$ \\ \hline
\end{tabular}
\caption{\label{tab:covfun} Supported covariance functions $c(u; \btheta)$ defined by their respective correlation functions $r(h)$, such that $c(u; \btheta) = \sigma^2 r(u/\rho)$. Compactly supported correlation functions are defined using the shorthand $[x]_{+}$, i.e., the positive part of $x$ and 0 otherwise.}
\end{table}

\begin{figure}
\centering
\includegraphics{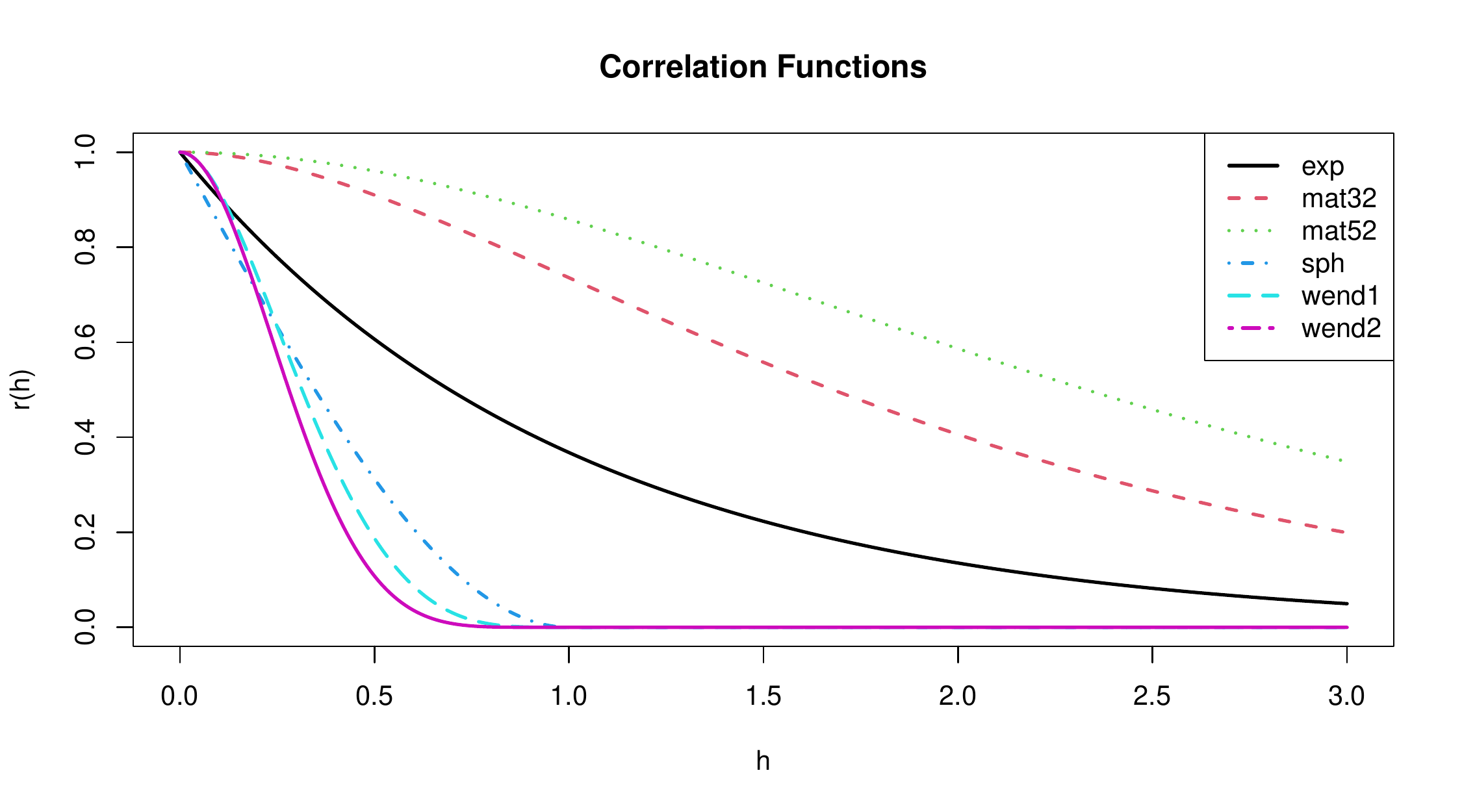}
\caption{\label{fig:cor} Correlation functions $r(h)$ as defined in Table~\ref{tab:covfun}.}
\end{figure}

\subsubsection{Profile Likelihood}

The parameter of interest $\bomega$ is of dimension $p + 2q + 1$. For numeric optimization, such a high dimensional parameter space tends to have numerical instabilities. In order to reduce the computational burden, one can concentrate the log likelihood and optimize on the profile likelihood. From~\eqref{eq:normalY}, we can see that $\Y$ has the form of a generalized linear model. The mean effect parameter $\bmu$ is a nuisance parameter, which can explicitly be defined by the covariance parameter~$\btheta$, i.e.:
\begin{align*}
  \bmu(\btheta) = \left(\X^\top \bSigma_{\Y}^{-1}(\btheta ) \X \right)^{-1} \X^\top  \bSigma_{\Y}^{-1}(\btheta ) \y.
\end{align*}
Therefore, the optimization can be expressed as only depending on the covariance parameter~$\btheta$ or, in other words, being expressed by an isometric profile of the full likelihood. The binary argument \code{profileLik} of \fct{SVC\_mle\_control} toggles if the optimization is to be computed on the profile likelihood, or not. 

\subsubsection{Initial Values and Boundary Conditions}

The numeric optimization -- both over the profile likelihood and the full likelihood -- requires providing initial values $\bomega^{(init)} \in \bOmega$. If not otherwise specified, we provide the following initial values. With respect to the mean parameters, we use the results from an ordinary least squares $\hat{\bmu}(\text{OLS})$ of the linear model $\y = \X\bmu + \bvarepsilon$. Concerning the covariance parameters, we use the median recorded distance between observations $\delta$ and the empirical variance of the response $\y$ denoted by $s_\y^2$. In particular, we set the initial values for the range to $\delta/4$ and for the variance of the Gaussian processes (including the nugget) to $s_\y^2/(q+1)$. Another possibility to set the initial values of the latent Gaussian processes is to first apply a GWR. With the estimated coefficients we are able to compute the semivariograms and to deduct the respective covariance parameters. Latter method is currently not implemented.

To ensure that the covariance matrix $\bSigma_\Y(\btheta)$ is well-defined and positive-definite, we require and check that $\sigma^2_k \geq 0, \rho_k > 0$, and $\tau^2 >0$ for all $k = 1, ..., q$. Additionally, the ``L-BFGS-B'' optimization by \citet{Byrd1995} allows lower bounds $\l \in \bOmega$ and upper bounds $\bfu \in \bOmega$ on the parameter of interest $\bomega \in \bOmega$, i.e., $\l \leq \bomega \leq \bfu$.  These bounds can benefit the stability of the numeric optimization, but have to be chosen carefully. For instance, if the range parameter $\rho_k$ of a covariance function is much larger than the diameter of the (observed) domain $D$, there is little to say about the corresponding Gaussian process. In fact, one could consider the coefficient to be non varying at all. On the other hand, if such upper bound is set too low, it hinders the optimization to obtain the best estimate. We can argue similarly for the lower bound. If a range parameter $\rho_k$ becomes too small, the estimated Gaussian process starts to behave like a nugget and an appropriately chosen lower bound $\l$ could prevent this. To summarize, we give the default initial values and boundaries for each type of parameter in Table~\ref{tab:liu}.

\begin{table}[t!]
\centering
\renewcommand{\arraystretch}{1.4}
\begin{tabular}{lrrr}
\hline
\textbf{Parameter type} & \textbf{Lower bound} $l_j$ &  \textbf{Initial value} $\omega^{(init)}_j $  & \textbf{Upper bound} $u_j$ \\ \hline
Range             &	$10^{-3}\delta$ & $\delta/4$        & $10\delta$ \\
Variance	        & $0$             &	$s^2_\y/(q+1)$    &	$10s^2_\y$ \\
Nugget	Variance  & $10^{-6}$       & $s^2_\y/(q+1)$    &	$10s^2_\y$ \\
Mean              & \code{-Inf}     &	$\hat{\mu}_j(\text{OLS})$ & \code{Inf} \\ \hline
\end{tabular}
\caption{\label{tab:liu} Default parameter settings, i.e., if no other arguments for \code{lower}, \code{init}, or \code{upper} of \fct{SVC\_mle\_control} are provided, respectively. The settings for the mean parameters are only required for an optimization over the full likelihood.}
\end{table}

\begin{Schunk}
\begin{Sinput}
R> # covariance parameter length: 2 GPs with each 2 parameters + nugget 
R> P <- 2*2+1
R> # setting bounds and initial values (overwriting the default values)
R> SVC_mle_control(
+    lower = c(rep(c(0.1, 0), 2), 1e-6), 
+    init = rep(1, P), 
+    upper = rep(Inf, P)
+  )
\end{Sinput}
\end{Schunk}

\subsubsection{Covariance Tapering}

Introduced by \citet{Furrer2006}, covariance tapering is a technique to tackle the ``big $n$ problem'', which arises from a large number of observations $n$ \citep{Heaton2019, Lasinio2013}. In particular, for $n>$ 10'000 computers struggle to calculate the determinant or solving a linear system of $\bSigma_{\Y}$ in a feasible time. Here, covariance tapering is a pragmatic approach which makes the evaluation of such calculations -- and therefore of the log likelihood \eqref{eq:ll} -- time efficient by inducing a sparse matrix structure. 

Therefore, the original covariance function~$c$ is multiplied with another, suitable covariance function~$c^\star$ with a sufficiently small support, i.e., $c^\star(u) = 0$ for all $u$ greater than some taper distance $\rho_\star>0$. The covariance function $c^{\star}$ has to be chosen in accordance with the original covariance function~$c$, see \citet{Furrer2006} for details. The tapered covariance function $c^{(tap)}(u) := c(u)c^{\star}(u)$ maintains most of its original characteristics, like the order of differentiability at the origin. Due to the small support of $c^{(tap)}(u)$, the induced covariance matrix becomes sparse and can be stored efficiently. Further, there exist optimized algorithms to take advantage of the sparse matrix structure, e.g., the Cholesky decomposition by \citet{NG1993} implemented in the \proglang{R} package \pkg{spam} \citep{R:spam}. The optimization of the likelihood is either executed on matrices without covariance tapering or on sparse matrices of class \class{spam}, where covariance tapering has been applied. Here, the \fct{SVC\_mle\_control} argument \code{tapering} triggers covariance tapering. By default, it is set to \code{NULL} and no covariance tapering is applied. If a positive scalar is provided, it defines the taper range $\rho_\star$ and therefore the sparsity structure of the covariance matrix.

\begin{Schunk}
\begin{Sinput}
R> # setting the taper range to distance 5 in the units of the locations
R> SVC_mle_control(tapering = 5)
\end{Sinput}
\end{Schunk}

\subsubsection{Parallelized Optimization}

The optimization of the log likelihood function \eqref{eq:ll} is implemented with the \proglang{R} function \fct{optim}, specifically the ``L-BFGS-B'' quasi-Newton gradient method \citep{Byrd1995}. In each iteration step~$\iota$, \fct{optim} sequentially evaluates the objective function $fn$ several times. Let $\bomega^{(\iota)} \in \bOmega$ be the current parameter value. First \fct{optim} evaluates $fn\bigl( \bomega^{(\iota)} \bigr)$ and then approximates its gradient $gr$ at $\bomega^{(\iota)}$ by evaluating $fn\bigl( \bomega^{(\iota)} + \bepsilon_p \bigr)$ for some $\bepsilon_p \in \IR^{|\bOmega|}, p = 1, ..., P$. The number of evaluations~$P$ to approximate the gradient generally increases with the dimension of the parameter space $\bOmega$, i.e., $|\bOmega|:=\dim\bOmega$. However, all of the above mentioned evaluations of $fn$ are independent of each other and we can take advantage of a parallel computing. It is implemented with the package \pkg{optimParallel} \citep{R:optimParallel}, where we only have to provide an initialized cluster as part of a \class{list} object in the argument \code{parallel}. In the documentation of \fct{SVC\_mle\_control} we give the following code to initialize parallel computing.
\begin{Schunk}
\begin{Sinput}
R> # Code for setting up parallel computing
R> require(parallel)
R> # exchange number of nodes (1) for detectCores()-1 or appropriate number
R> cl <- makeCluster(1)
R> clusterEvalQ(
+    cl = cl,
+    {
+      library(spam)
+      library(varycoef)
+    })
R> # use this list for parallel argument in SVC_mle_control
R> parallel.control <- list(cl = cl, forward = TRUE, loginfo = TRUE)
R> # SVC_mle goes here ...
R> # DO NOT FORGET TO STOP THE CLUSTER!
R> stopCluster(cl); rm(cl)
\end{Sinput}
\end{Schunk}

\subsection{Maximum Likelihood Estimation}

Once the control parameters have been set, the subsequent maximum likelihood estimation is straight forward. The corresponding function is \fct{SVC\_mle}, where one has to provide the before mentioned control settings from \fct{SVC\_mle\_control} and data. In particular, the function requires a numeric vector \code{y} of length \code{n} as the response, the fixed effects data \code{X} as a matrix with dimension (\code{n, p}) and the locations \code{locs}. The latter should be provided as a matrix of dimension (\code{n, d}), where \code{d} is the dimension of the domain $D$. In contrast to most methodologies for SVC models, the domain $D$ does not have to be a subset of $\mathbb{R}^2$, i.e., with $d = 2$. If the matrix \code{W} is not defined, all covariates provided for the fixed effects will be used to model random effects, i.e., SVCs, too. Otherwise, the matrix \code{W} of dimension (\code{n, q}) explicitly defines the covariates $\w^{(k)}$. The estimated parameter of interest by maximizing the likelihood is denoted $\hat{\bomega}(\text{MLE})$. We give two examples. The first one is the sampled data set with $d = 1$ from above. The second one is a real data set of approximately 25'000 observations with $d = 2$. 

\subsubsection{Simple SVC Model} We use the data generated with \fct{sample\_fullSVC} and displayed in Figure~\ref{fig:simpleSVCmodel}, i.e., we take the data contained in \code{SVCdata}. We optimize over the profile likelihood and set the covariance function to be a Mat\'ern with $\nu = 3/2$. All other control parameters are the default ones.
\begin{Schunk}
\begin{Sinput}
R> # set control parameters
R> crtl <- SVC_mle_control(cov.name = "mat32", profileLik = TRUE)
R> # Run MLE (takes approximately one minute)
R> fit <- with(SVCdata, SVC_mle(y = y, X = X, locs = locs, control = crtl))
R> summary(fit)
\end{Sinput}
\begin{Soutput}
Call:
SVC_mle.default(y = y, X = X, locs = locs, control = crtl)

Fitting a GP-based SVC model with 2 fixed effect(s) and 2 SVC(s)
using 300 observations at 300 different locations / coordinates.

Residuals:
     Min.    1st Qu.     Median    3rd Qu.       Max.  
-1.334351  -0.318621   0.003409   0.315742   1.272693  

Residual standard error: 0.489
Multiple R-squared: 0.9599, BIC: 657.7


Coefficients of fixed effect(s):
     Estimate Std. Error Z value Pr(>|Z|)    
Var1   1.1799     0.3820   3.089  0.00201 ** 
Var2   2.2225     0.6675   3.329  0.00087 ***
---
Signif. codes:  0 '***' 0.001 '**' 0.01 '*' 0.05 '.' 0.1 ' ' 1


Covariance parameters of the SVC(s):
           Estimate Std. Error W value Pr(>W)  
SVC1.range  0.32576    0.07048      NA     NA  
SVC1.var    1.22053    0.47731   6.539 0.0106 *
SVC2.range  0.85962    0.26378      NA     NA  
SVC2.var    1.63277    1.02824   2.522 0.1123  
nugget.var  0.29060    0.02708      NA     NA  
---
Signif. codes:  0 '***' 0.001 '**' 0.01 '*' 0.05 '.' 0.1 ' ' 1

The covariance parameters were estimated using 
Matern (nu = 3/2) covariance functions.
No covariance tapering applied.


MLE:
The MLE terminated after 69 function evaluations with convergence code 0
(0 meaning that the optimization was succesful).
The final profile log likelihood value is -317.4.
\end{Soutput}
\end{Schunk}
The summary output of the function \fct{SVC\_mle} provides an overview of the data and model, the estimated parameters for the fixed and random effects, as well as the summary of the optimization. If possible, the estimates' standard errors are approximated using the Hessian of the optimization. Further, we use a $Z$ test on the fixed effects ($H_0$: $\mu_j = 0$) and a Wald test on the Gaussian process variances ($H_0$: $\sigma_k^2 = 0$). The other covariance parameters, i.e., the ranges $\rho_k$ and the nugget variance $\tau^2$, are defined to be strictly positive. Therefore no Wald test is conducted and the corresponding test statistics and $p$-values for these parameters are always set to \code{NA}.

\subsubsection{Lucas County House Price Data}

A real, larger data set is given by the Lucas County (OH) from the \proglang{R} package \pkg{spData} \citep{R:spData}. This data set is available as a \class{data.frame} in \pkg{varycoef} using \code{data("house")}. We use a subset of covariates in our model and give a brief overview thereof in Table~\ref{tab:House}. The continuous covariates were transformed to account skewness and to increase numeric stability. We denote these transformed variables on the covariates using the prefix \code{Z} for a standardization and \code{l} for a logarithmic transformation using $\log(x + 1)$. The SVC model also contains the transformed year of construction as a quadratic effect to account for a potential vintage effect (see~\citealp{JAD2020:2}). This results in a model with six varying coefficients and 20 fixed effects including the mean.

\begin{figure}
\centering
\includegraphics{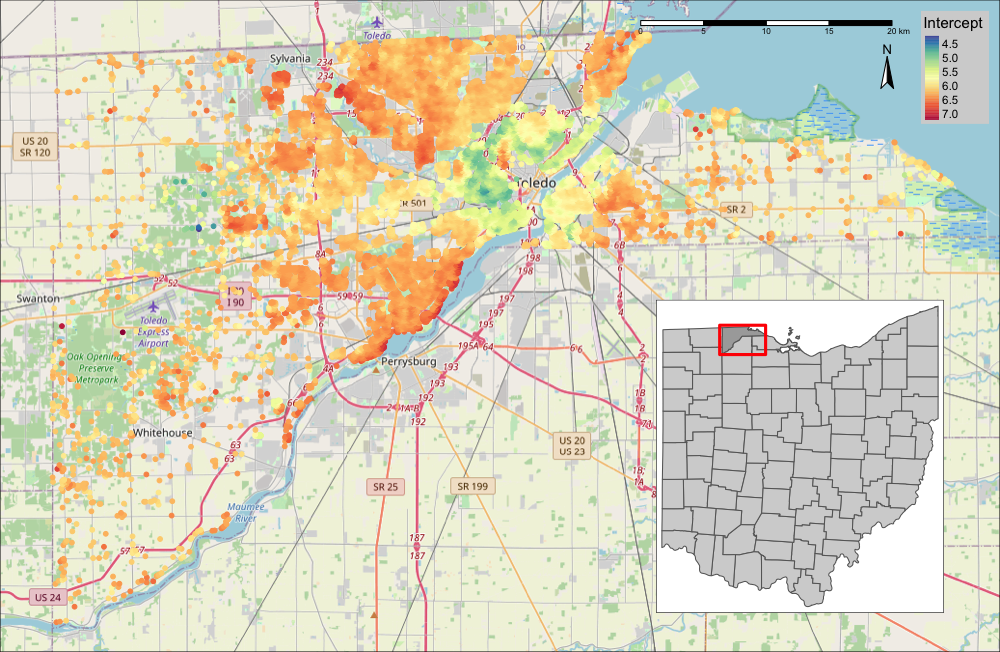}
 \vspace{1cm}
 \includegraphics{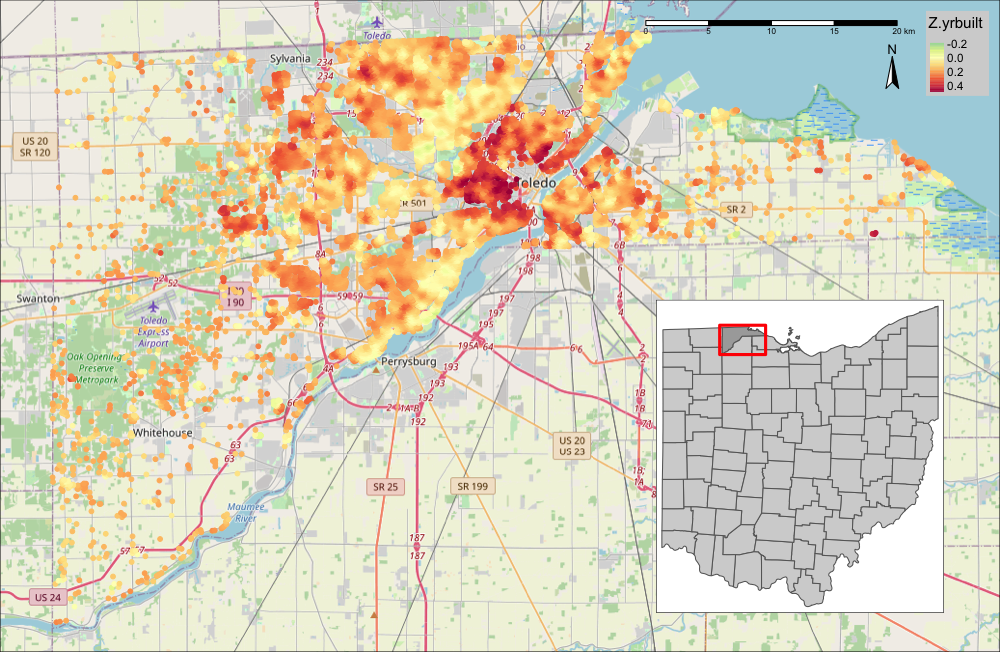}
\caption{\label{fig:SVC-LC} Two estimated spatially varying coefficients for the intercept (upper panel) and the standardized year of construction (lower panel). The inset maps shows the State of Ohio with Lucas county filled in dark grey.}
\end{figure}

\begin{table}[!t]
\centering
\begin{tabular}{lllrrrr}
  \multicolumn{7}{l}{\emph{(a) Continuous Variables}} \\ \hline & \textbf{Variable} &	\textbf{Description}& \multicolumn{4}{c}{\textbf{Summary Statistics}} \\ \cmidrule(lr){4-7} &  &  & Min. & Mean & SD & Max. \\ 
  \hline
\multirow{1}{*}{\rotatebox[origin=c]{90}{$y$}} & \texttt{price} & transaction price in USD & 2000 & 79018 & 59655 & 875000 \\ 
   \hline
\multirow{4}{*}{\rotatebox[origin=c]{90}{$x$}} & \texttt{yrbuilt} & building year & 1835 & 1945 & 28 & 1998 \\ 
   & \texttt{TLA} & total living area in square feet & 120 & 1462 & 613 & 7616 \\ 
   & \texttt{lotsize} & lot size in square feet & 702 & 13332 & 28941 & 429100 \\ 
   & \texttt{garagesqft} & garage area in square feet & 0 & 370 & 208 & 5755 \\ 
   \hline
\multirow{2}{*}{\rotatebox[origin=c]{90}{$s$}} & \texttt{long} & longitude in meters & 484575 & 508144 & 6948 & 538364 \\ 
   & \texttt{lat} & latitude in meters & 195270 & 221710 & 5095 & 229836 \\ 
   \hline \multicolumn{7}{l}{} \\ \multicolumn{7}{l}{\emph{(b) Factor Variables}} \\ \hline & \textbf{Variable} &	\textbf{Levels}& \multicolumn{4}{c}{\textbf{Frequency}} \\ \cmidrule(lr){4-7}  & & & \multicolumn{2}{r}{Absolute} & \multicolumn{2}{r}{Relative [in \%]} \\ \hline
\multirow{19}{*}{\rotatebox[origin=c]{90}{$x$}} & \texttt{stories} & \texttt{one} &  & 12954 &  & 51 \\ 
   &  & \texttt{bilevel} &  & 509 &  & 2 \\ 
   &  & \texttt{multilvl} &  & 723 &  & 3 \\ 
   &  & \texttt{one+half} &  & 3125 &  & 12 \\ 
   &  & \texttt{two} &  & 8042 &  & 32 \\ 
   &  & \texttt{two+half} &  & 2 &  & 0 \\ 
   &  & \texttt{three} &  & 2 &  & 0 \\ 
   \cmidrule(lr){2-7} & \texttt{wall} & \texttt{stucdrvt} &  & 204 &  & 1 \\ 
   &  & \texttt{ccbtile} &  & 129 &  & 1 \\ 
   &  & \texttt{metlvnyl} &  & 4235 &  & 17 \\ 
   &  & \texttt{brick} &  & 3633 &  & 14 \\ 
   &  & \texttt{stone} &  & 86 &  & 0 \\ 
   &  & \texttt{wood} &  & 11174 &  & 44 \\ 
   &  & \texttt{partbrk} &  & 5896 &  & 23 \\ 
   \cmidrule(lr){2-7} & \texttt{garage} & \texttt{no garage} &  & 3488 &  & 14 \\ 
   &  & \texttt{basement} &  & 78 &  & 0 \\ 
   &  & \texttt{attached} &  & 9018 &  & 36 \\ 
   &  & \texttt{detached} &  & 12555 &  & 50 \\ 
   &  & \texttt{carport} &  & 218 &  & 1 \\ 
   \hline
\end{tabular}
\caption{Description and summary statistics, i.e., the minimum, mean, standard deviation, and maximum of continuous and frequencies of factor variables in Lucas County house data. The respective types of data are given in the first column, where $y$ denotes the response, $x$ denotes explanatory variables, and $s$ denotes the observation location in a $d=2$ dimensional domain, i.e., Northing and Easting in Ohio North coordinate reference system (\code{epsg:2834}). The first listed levels of the factors are the reference levels.} 
\label{tab:House}
\end{table}
Due to the large observation size ($n =$~ 25'353 not counting the observations with 2.5 and 3 stories) and the relatively large number of SVCs ($q = 6$), we apply covariance tapering with a taper distance of 1~kilometer, optimize over the profile likelihood, and use parallel computing. The parameter estimates are given in Table~\ref{tab:Est}. In Figure~\ref{fig:SVC-LC} we present two estimated SVC with the strongest spatial structure combined with their fixed effect, i.e., $\hat{\mu}_1 + \hat{\bfeta}_1(\cdot)$ and $\hat{\mu}_2 + \hat{\bfeta}_2(\cdot)$ respectively corresponding to the intercept and the standardized year of construction. The code for the SVC model estimation is given in the appendix (c.f.~Section~\ref{subsec:SVCModel}).

Some key insights of the model can be immediately extracted. For instance, the downtown area of Toledo has the lowest mean pricing, while house prices are highest along the shore line of the Maumee River close to Perrysburg. Northwest and West of Toledo are a couple of local, high pricing areas. These features can also be obtained from classical geo-statistical models where we model a spatially varying intercept. For the effect of the year of construction \code{yrbuilt} we observe some interesting behavior. For a majority of locations the coefficient is clearly positive. The strongest, positive effect is present at the downtown area. Over all, we interpret these results as high desirability of newly built houses. However, in the suburbs of Toledo and along the Maumee River, we clearly see a deviation of this behavior as there are some locations where the \code{yrbuilt} coefficient is close to zero, or even negative. This hints at a vintage effect being present. We refer to \citet{JAD2020:2} where an similar analysis for single family houses in the Canton of Zurich (Switzerland) is conducted. The remaining estimated spatially varying coefficients are given in the appendix (c.f.~Section~\ref{subsec:remainSVC}).

%
\begin{table}[!t]
\centering
\begin{tabular}{lrrrrrr}
  \hline \textbf{Variable} & \multicolumn{2}{c}{\textbf{Mean} $\hat{\mu}_j$}	&\multicolumn{2}{c}{\textbf{Range} $\hat{\rho}_k$} & \multicolumn{2}{c}{\textbf{Variance} $\hat{\sigma}^2_k$} \\ \cmidrule(lr){2-3} \cmidrule(lr){4-5} \cmidrule(lr){6-7}  & Est. & SE & Est. & SE & Est. & SE  \\ 
  \hline
\texttt{(Intercept)} & 6.130 & 0.064 & 269.202 & 225.676 & 0.09408 & 0.00343 \\ 
  \texttt{Z.yrbuilt} & 0.163 & 0.007 & 50.795 & 68.081 & 0.01795 & 0.00142 \\ 
  \texttt{Z.yrbuilt.sq} & $-$0.026 & 0.004 & 0.011 &  & 0.01221 & 0.00037 \\ 
  \texttt{l.TLA} & 0.499 & 0.007 & 101.474 &  & 0.00000 & 0.00000 \\ 
  \texttt{l.lotsize} & 0.127 & 0.004 & 101.472 &  & 0.00000 & 0.00000 \\ 
  \texttt{l.garagesqft} & 0.075 & 0.005 & 101.397 & 219.921 & 0.00010 & 0.00005 \\ 
  \texttt{storiesbilevel} & $-$0.062 & 0.013 &  &  &  &  \\ 
  \texttt{storiesmultilvl} & $-$0.026 & 0.010 &  &  &  &  \\ 
  \texttt{storiesone+half} & 0.026 & 0.006 &  &  &  &  \\ 
  \texttt{storiestwo} & 0.056 & 0.005 &  &  &  &  \\ 
  \texttt{wallccbtile} & $-$0.171 & 0.025 &  &  &  &  \\ 
  \texttt{wallmetlvnyl} & 0.037 & 0.016 &  &  &  &  \\ 
  \texttt{wallbrick} & 0.063 & 0.016 &  &  &  &  \\ 
  \texttt{wallstone} & 0.016 & 0.027 &  &  &  &  \\ 
  \texttt{wallwood} & $-$0.007 & 0.016 &  &  &  &  \\ 
  \texttt{wallpartbrk} & 0.033 & 0.016 &  &  &  &  \\ 
  \texttt{garagebasement} & $-$0.256 & 0.041 &  &  &  &  \\ 
  \texttt{garageattached} & $-$0.271 & 0.032 &  &  &  &  \\ 
  \texttt{garagedetached} & $-$0.300 & 0.032 &  &  &  &  \\ 
  \texttt{garagecarport} & $-$0.345 & 0.033 &  &  &  &  \\ 
  Nugget &  &  &  &  & 0.03143 & 0.00046 \\ 
   \hline
\end{tabular}
\caption{Parameter estimates (Est.) and corresponding standard errors (SE) of GP-based SVC model for Lucas county data. If no corresponding standard error is given, then it could not be retrieved from the Hessian.} 
\label{tab:Est}
\end{table}%

\subsection{Prediction}

\begin{figure}
\centering
\includegraphics{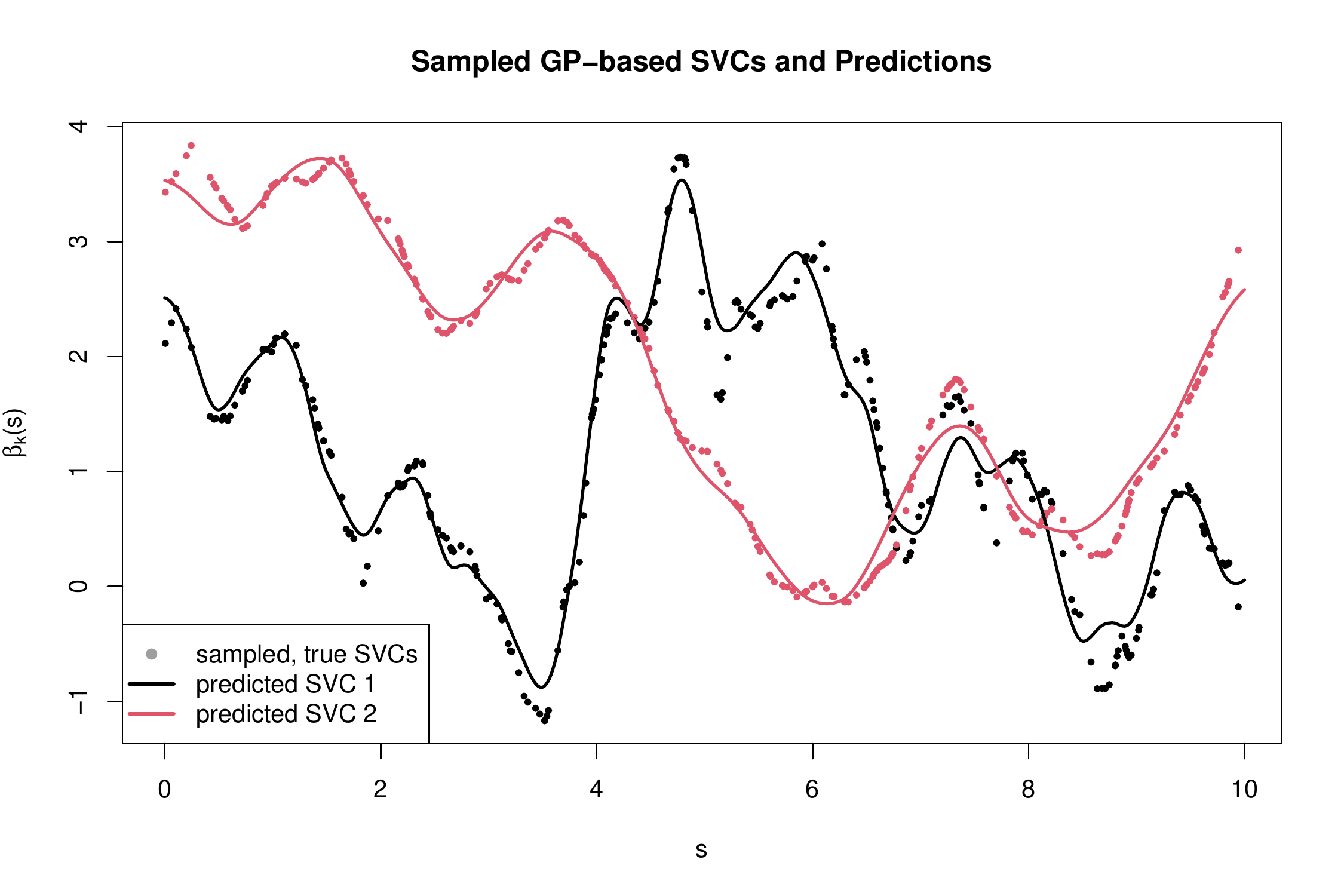}
\caption{\label{fig:SVCexample} Sampled, true SVCs from \code{SVCdata} (dots) and predicted SVCs (lines) using MLE of GP-based SVC model.}
\end{figure}

Once the parameter of interest $\hat{\bomega}(\text{MLE})$ has been found, we can calculate (spatial) predictions using the empirical best linear unbiased predictor (EBLUP). In \pkg{varycoef}, a \fct{predict} method for the class \class{SVC\_mle}, i.e., the model output of \fct{SVC\_mle}, has been implemented. The argument \code{newlocs} takes the new locations to spatially predict the random effects. If arguments \code{newX} and \code{newW} are provided, the response and predicitive variance are also calculated. Note that the predicted random effects only contain $\hat{\bfeta}_k$, i.e., the zero-mean Gaussian processes. If a corresponding mean effect is associated to the same covariate, it can be retrieved by the \fct{coef} method and added to the predicted Gaussian process, similar to Figure~\ref{fig:SVC-LC}. Finally, we visualize the true and predicted varying coefficients by our MLE approach in Figure~\ref{fig:SVCexample}.
\begin{Schunk}
\begin{Sinput}
R> # Predict SVCs on whole interval
R> newlocs <- seq(0, 10, by = 0.01)
R> SVCpred <- predict(fit, newlocs = newlocs)
R> # Combine random effects from GPs
R> SVCpredbeta <- as.matrix(SVCpred[, 1:2]) + 
+    # and fix effects
+    matrix(coef(fit), nrow = length(newlocs), ncol = 2, byrow = TRUE)
R> head(SVCpredbeta)
\end{Sinput}
\begin{Soutput}
        SVC_1    SVC_2
[1,] 2.509802 3.533227
[2,] 2.506545 3.530252
[3,] 2.501862 3.527029
[4,] 2.495821 3.523557
[5,] 2.488497 3.519835
[6,] 2.479956 3.515861
\end{Soutput}
\end{Schunk}

\section{Variable Selection}\label{sec:PMLE}

\subsection{Introduction}

Due to the flexible nature of a GP-based SVC models, some natural questions arise when defining the model or interpreting the results of an estimated model: \emph{Which covariates should be defined with spatially varying coefficients? Is a constant coefficient sufficient? Did the estimated model overfit the data due to its high flexibility?} To address these questions, we introduced a joint variable selection method for the fixed and random effects of the GP-based SVC model \citep{JAD2021}.

\subsection{Optimization of the Penalized Likelihood}

The penalized likelihood is defined by the likelihood $\ell$ \eqref{eq:ll} and $L_1$ penalties \citep{Tibshirani1996} on the fixed effects $\mu_j$ and the variance $\sigma_k^2$, i.e., 
\begin{align}
  p\ell(\bomega) = \ell(\bomega) + n \sum_{j = 1}^p \lambda_j |\mu_j| + \sum_{k = 1}^q \lambda_{p+k} |\sigma^2_k|  \label{eq:pll}
\end{align}
The optimization problem is related to variable selection of linear mixed models (for an overview, see \citealp{Mueller2013}) and, in particular, the works of \citet{Bondell2010} and \citet{Ibrahim2011}. We assume that the shrinkage parameters of the penalized likelihood \eqref{eq:pll} are defined like in an adaptive Lasso \citep{Zou2006}. However, we account for their inherit difference as the parameters shrink the fixed and random effects, respectively. For the unknown shrinkage parameters $\bigl(\lambda_\bmu, \lambda_\btheta\bigr) \in \bigl(\IR_{>0} \bigr)^2$, we have: 
\begin{align}
  \lambda_{j} := \frac{\lambda_\bmu}{\hat{\mu}_j}, \quad \lambda_{p+k} := \frac{\lambda_\btheta}{\hat{\sigma}^2_k}.
\end{align}
For now, we assume that $\bigl(\lambda_\bmu, \lambda_\btheta\bigr)$ are fixed and known. The objective function of the optimization is defined as $-2p\ell(\bomega)$. Its optimization is achieved by a coordinate descent where we cyclically iterate between the optimization of the fixed effects $\bmu$ and the covariance parameters~$\btheta$, i.e.,
\begin{align*}
  \bmu^{(t+1)} &= \argmin_{\bmu \in \IR^p} -2p\ell(\bmu | \btheta^{(t)}), \\
  \btheta^{(t+1)} &= \argmin_{\btheta \in \bTheta} -2p\ell(\btheta | \bmu^{(t+1)}),
\end{align*}
for $t\geq 0$, where the initial value is given by $\btheta^{(0)} = \hat{\btheta}(\text{MLE})$. While the first step is implemented by the \pkg{glmnet} package \citep{Friedman2010}, the second step requires more effort. We refer to \citet{JAD2021} for more details. Under convergence, the coordinate descent approach returns the penalized maximum likelihood estimates denoted $\hat{\bomega}(\text{PMLE})$.

\subsection{Selection of Shrinkage Parameters}

In the last section, we assumed the shrinkage parameters $\bigl(\lambda_\bmu, \lambda_\btheta\bigr)$ to be known. Here, we focus on their selection by optimizing an information criterion. For some shrinkage parameters $\bigl(\lambda_\bmu, \lambda_\btheta\bigr)$, we call the coordinate descent which computes $\hat{\bomega}\bigl(\lambda_\bmu, \lambda_\btheta\bigr)$ for respective shrinkage parameters, which we abbreviate with $\hat{\bomega}_{\blambda}$. For the estimated model with parameters $\hat{\bomega}_{\blambda}$ we can calculate an information criterion.

Currently, there are two information criteria implemented in \pkg{varycoef}: a conditional Akaike (cAIC) introduced by \citet{Vaida2005} and a Bayesian (BIC) information criterion. For parameter estimates $\hat{\bomega}_{\blambda}$, both information criteria are defined as the sum of the goodness of fit given with $-2\ell\bigl(\hat{\bomega}_{\blambda}\bigr)$ (also called deviance) and a specific model complexity penalty $\alpha\bigl(\hat{\bomega}_{\blambda}\bigr)$. For the cAIC's penalty we require the effective degrees of freedom $df(\cdot)$ which are defined as the trace of the hat matrix $\H$:
\begin{align*}
  df\bigl(\hat{\bomega}_{\blambda}\bigr) &= \trace \H\bigl(\hat{\btheta}_{\blambda}\bigr)  \\
  &= \hat{\tau}_\blambda^2\trace \left[\bigl( \X^\top \bSigma^{-1} \X \bigr)^{-1}  \X^\top  \bSigma^{-1}\bSigma^{-1}\X\right] + n - \hat{\tau}^2_\blambda \trace \left[\bSigma^{-1}\right],
\end{align*}
where $\bSigma^{-1} = \left[\bSigma \bigl( \hat{\btheta}_{\blambda} \bigr)\right]^{-1}$. The respective definitions of $\alpha\bigl(\hat{\bomega}_{\blambda}\bigr)$ are given in Table~\ref{tab:IC}. 

\begin{table}[t!]
\centering
\renewcommand{\arraystretch}{1.7}
\begin{tabular}{lll}
\hline
\code{IC.type}   & \textbf{Name}  & \textbf{Model complexity penalty} $\alpha\bigl(\hat{\bomega}_{\blambda}\bigr)$ \\ \hline
\code{"cAIC_VB"} & cond.\ Akaike IC & $\frac{2n}{n-p-2}\bigl(df(\hat{\bomega}_\blambda) + 1 - \frac{df(\hat{\bomega}_\blambda) - p}{n-p}\bigr)$ \\
\code{"BIC"}     & Bayesian IC & $log(n)\bigl( \|\bmu_\blambda\|_0 + \|\sigma^2_\blambda\|_0\bigr)$ \\ \hline
\end{tabular}
\caption{\label{tab:IC} Supported information criterion. The provided estimates $\hat{\bomega}_\blambda$ are obtained from a PMLE depending on the shrinkage parameters $\lambda_\bmu, \lambda_\btheta$, c.f.~Figure~\ref{fig:PMLEscheme}. The norm $\|\cdot\|_0$ is the number of non-zero elements.}
\end{table}

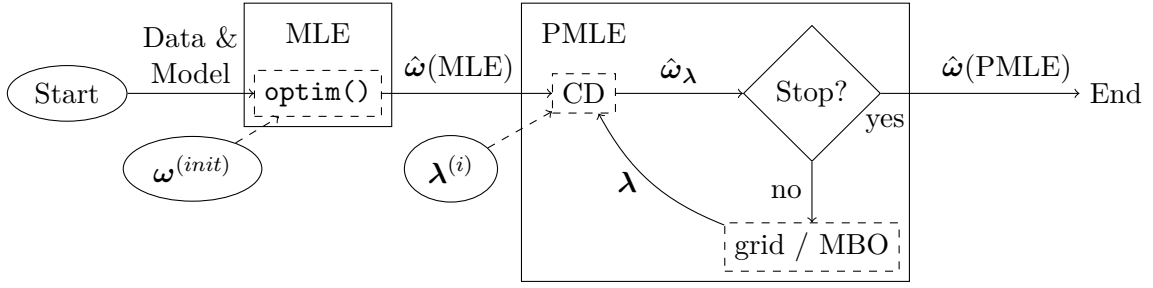
\begin{figure}
  \centering
		\begin{tikzpicture}[
  	node distance={35mm}, 
  	main/.style = {draw, rectangle},
  	sub/.style={draw, dashed, rectangle}
  ] 
	\node[ellipse, draw] (start) {Start};
	\node[sub, right of = start, node distance = 33mm] (optim) {\fct{optim}};		
	\node[above of = optim, node distance = 8mm] (mle) {MLE};
	\node[rectangle, draw, fit={(mle) (optim)}] (MLEall) {};
	\node[sub, right of = optim] (cd) {CD};	
	\node[above of = cd, node distance = 8mm] (pmle) {PMLE};
	\node[diamond, draw, right of = cd, node distance = 30mm] (conv) {Stop?};
	\node[sub, node distance = 8mm, below=of conv.south] (method) {grid / MBO};	
	\node[rectangle, draw, fit={(method) (conv) (pmle) (cd)}] (PMLEall) {};	
	\node[right of = conv, node distance = 40mm] (end) {End};
	\node[ellipse, draw, yshift = -10mm] (initMLE) at ($(start)!0.5!(optim)$) {$\bomega^{(init)}$};
	\node[ellipse, draw, yshift = -10mm] (initPMLE) at ($(optim)!0.5!(cd)$) {$\blambda^{(i)}$};

	\path[dashed, ->] (initMLE) edge (optim);
	\path[dashed, ->] (initPMLE) edge (cd);
	\draw[->] (start) -- node[midway, above, sloped, pos=0.5, align=center]{Data \& \\ Model} (optim);
	\draw[->] (optim) -- node[midway, above, sloped, align=center, xshift = -1mm]{$\hat{\bomega}(\MLE)$} (cd);
	\draw[->] (cd) -- node[midway, above, sloped, pos=0.5, align=center]{$\hat{\bomega}_\blambda$} (conv);
	\draw[->] (conv) -- node[near end, above, sloped, align=center, xshift = -3mm]{$\hat{\bomega}(\textnormal{PMLE})$} node[near start, below, xshift = -6mm, yshift = -1.5mm]{yes}(end);
	\draw[->] (conv) -- node[near end, rotate=90, left, sloped, align=center, yshift = 2mm]{no} (method);
	\path[->] (method) edge [bend left = 20] node[xshift=-3mm] {$\blambda$} (cd);  

  \end{tikzpicture} 
\caption{Variable selection scheme. The inputs are given by ellipses. Optional inputs are indicated by dashed arrows. The optimization routines are given in solid rectangles, where the algorithmic sub-routines are given in dashed rectangles, i.e., the quasi-Newton gradient method implemented in the \fct{optim} function, the coordinate descent (CD) algorithm, and the grid or MBO method. The stopping criterion within the PMLE depends on the optimization method. Latter method then provides the next $\blambda$ to estimate $\hat{\bomega}_\blambda$ by a coordinate descent.} \label{fig:PMLEscheme}
\end{figure}

We provide an overview of the variable selection scheme in Figure~\ref{fig:PMLEscheme}. Starting with the data and some model specification, we receive a first estimate via MLE, in particular using \fct{optim}. The estimate $\hat{\bomega}(\MLE)$ is then used in the PMLE. Hence, for a given shrinkage parameter $\blambda$ an estimate $\hat{\omega}_\blambda$ is returned. For such estimate we can calculate the respective information criterion. The whole procedure beginning with an input of a shrinkage parameter $\blambda$ over the PML-estimate $\hat{\omega}_\blambda$ to an information criterion output can be expressed as a function $IC(\cdot)$. Therefore, an information criterion optimization over corresponding objective function $IC(\cdot)$ selects a shrinkage parameter by minimizing $IC(\cdot)$. The back and forth between coordinate descent for a penalized likelihood optimization and the selection of the next shrinkage parameter is also visible in Figure~\ref{fig:PMLEscheme}. If a required stopping condition is met, the final PML-estimate is returned. Keep in mind that the evaluation of $IC(\cdot)$ is computationally expensive and complex. The information criterion optimization is therefore an optimization of an expensive (black-box) objective function over two parameters. We offer two methods on how to optimize the function:
\begin{itemize}
  \item The first method is a brute force grid approach by providing pairs of shrinkage parameters and computing it for all given combinations. Once the information criterion has been evaluated for each shrinkage parameter, the stopping condition is met. \\
  \item The second method is a much more sophisticated model-based optimization (MBO) which we briefly describe in the next section. 
\end{itemize}
The PMLE has its respective control function named \fct{SVC\_selection\_control} to set all the control parameters. In the data example discussed in Section~\ref{sec:uschange} as well as the Appendix~\ref{app:uschange} these control parameters are addressed.

\subsubsection{Model-based Optimization}

Model-based optimization \citep{Jones2001, Koch2012, Horn2016} offers compelling means to find a minimum of the objective function $IC(\cdot)$ by using a so called surrogate model. It relies on $n_\text{init}$ initial values $\blambda^{(1)}, ..., \blambda^{(n_\text{init})}$ that span the predefined parameter space $\Lambda$. By evaluating $IC$ for these values, we receive $n_{init}$ tuples $\bigl(\blambda^{(i)},\xi^{(i)}\bigr)$ with $\xi^{(i)} := IC(\blambda^{(i)})$, which we use to krige. More specifically, we assume a Gaussian surrogate model with constant mean and a Mat\'ern covariance function of smoothness $\nu = 3/2$ and estimated covariance parameters, c.f. equation~\eqref{eq:matern} and Table~\ref{tab:covfun}. In this case, the random variable $\Xi(\blambda)$ expressing the distribution at $\blambda$ conditional on the $n_\text{init}$ tuples is given by a normal distribution $\Xi(\blambda)\sim \cN \left(\widehat{\mu}(\blambda), \widehat{s}^2(\blambda) \right)$. Here, $\widehat{\mu}(\blambda)$ and $\widehat{s}^2(\blambda)$ are the kriging surface and kriging variance, respectively, for which we use the plug-in estimates of the surrogate model. The parameters of the surrogate model are then iteratively updated for $n_\text{iter}$ steps. In each step $\iota = 1, ..., n_\text{iter}$:
\begin{enumerate}
  \item Define the current information criterion minimum $\xi_\text{min} := \min \{ \xi^{(1)}, ..., \xi^{(n_\text{init}+\iota-1)}\}$.
  \item Compute an infill criterion from the current posterior distribution $\Xi(\blambda)$. We use the expected improvement (EI, see equation \eqref{eq:EI} below) infill criterion which can be expressed analytically for a Gaussian process surrogate model (see equation \eqref{eq:EI-GP} below):
  \begin{align}\label{eq:EI}
	\textnormal{EI}(\blambda) &= E_{\Xi}\left(\max \{ \xi_\text{min} - \Xi(\blambda), 0 \} \right) \\
	&= \begin{cases}
	  \bigl(\xi_\text{min} - \widehat{\mu}(\blambda) \bigr) \Phi\left( \frac{\xi_\text{min} - \widehat{\mu}(\blambda)}{\widehat{s}(\blambda)} \right) + \widehat{s}(\blambda) \phi\left( \frac{\xi_\text{min} - \widehat{\mu}(\blambda)}{\widehat{s}(\blambda)} \right), &\quad \textnormal{if } \widehat{s}(\blambda) > 0, \\
	  0, &\quad \textnormal{if } \widehat{s}(\blambda) = 0. 
	\end{cases} \label{eq:EI-GP}
\end{align}
  In the equation above $\Phi$ and $\phi$ denote the cumulative distribution function and the probability density function of the standard normal distribution, respectively. The next, best shrinkage parameter $\blambda^{(n_\text{init} + \iota)}$ is found by maximizing  \eqref{eq:EI-GP}.
  \item Evaluate the information criterion at respective location, i.e., $\xi^{(n_\text{init} + \iota)} := IC(\blambda^{(n_\text{init} + \iota)})$, and add the tuple $\bigl(\xi^{(n_\text{init} + \iota)}, \blambda^{(n_\text{init} + \iota)}\bigr)$ to the existing set of tuples.
  \item Update the surrogate model's parameters and the distribution of $\Xi(\blambda)$. 
\end{enumerate}
The $n_\text{init}$ initial shrinkage parameters $\blambda^{(i)}$ are drawn as Latin hypercube sample (LHS) from a predefined subset of the parameter space $\Lambda$ using the \proglang{R} packages \pkg{lhs} \citep{R:lhs} and \pkg{ParamHelpers} \citep{R:ParamHelpers}. The surrogate model is defined using the \proglang{R} package \pkg{mlr} \citep{Bischl2016}. The MBO is implemented with the \proglang{R} package \pkg{mlrMBO} \citep{R:mlrMBO}.

\subsection{A Final Example: Growth Rates in the USA}\label{sec:uschange}

We showcase the variable selection with a small time series data set, the \code{uschange} data set in the \proglang{R} package \pkg{fpp2} \citep{Hyndman2018}. It contains the quarterly percentage changes of personal consumption (\code{Consumption}) and personal disposable income (\code{Income}) as well as the US production (\code{Production}), savings (\code{Savings}), and the unemployment (\code{Unemployment}) rates from Q1~1970 to Q3~2016.
\begin{Schunk}
\begin{Sinput}
R> data("uschange", package = "fpp2")
R> # divide times series into data and time points of observation
R> train_dat <- as.data.frame(uschange); train_t <- as.numeric(time(uschange))
\end{Sinput}
\end{Schunk}
\begin{figure}
\centering
\includegraphics{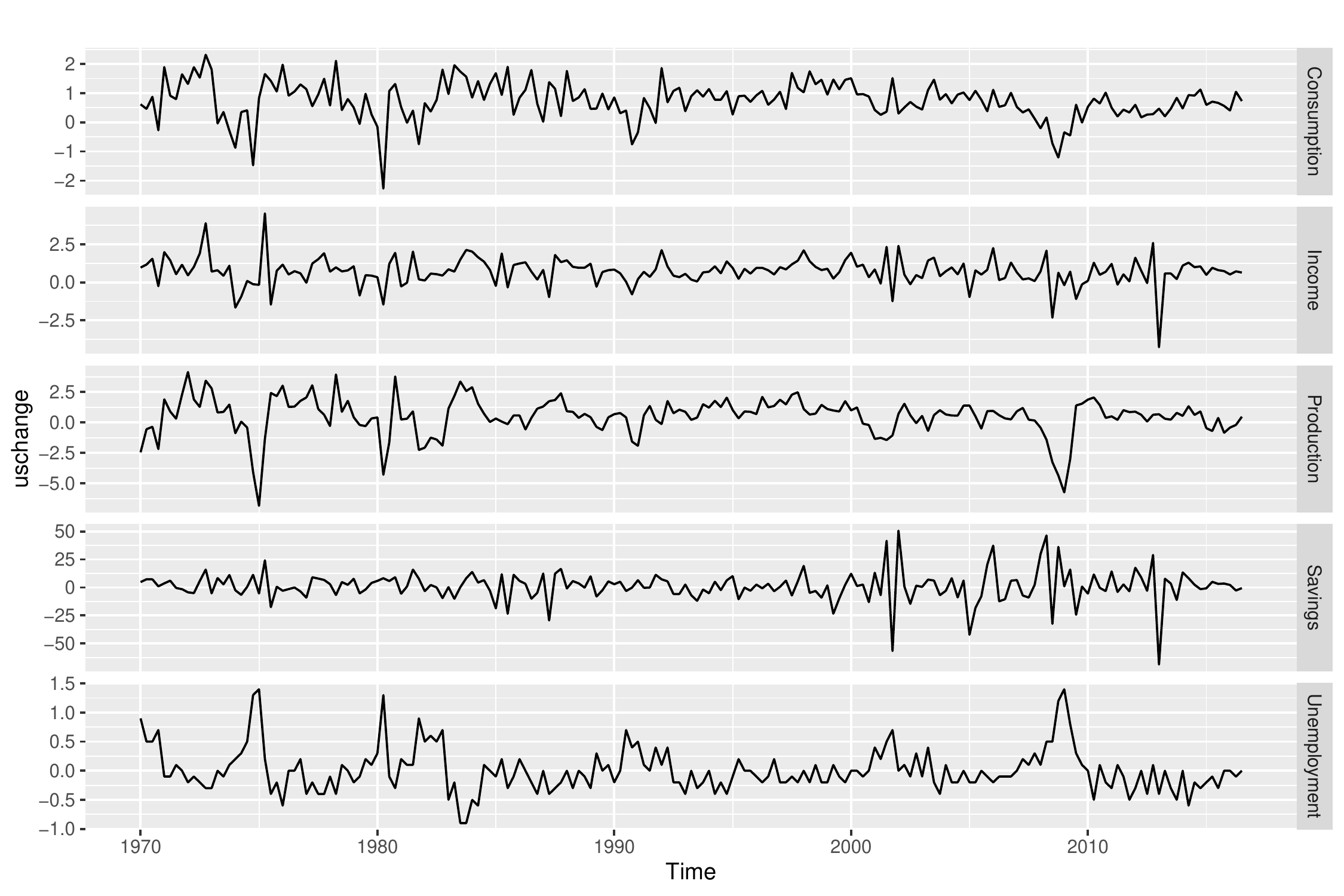}
\caption{\label{fig:USCtime} Time series of \code{uschange} data.}
\end{figure}
\begin{figure}
\centering
\includegraphics{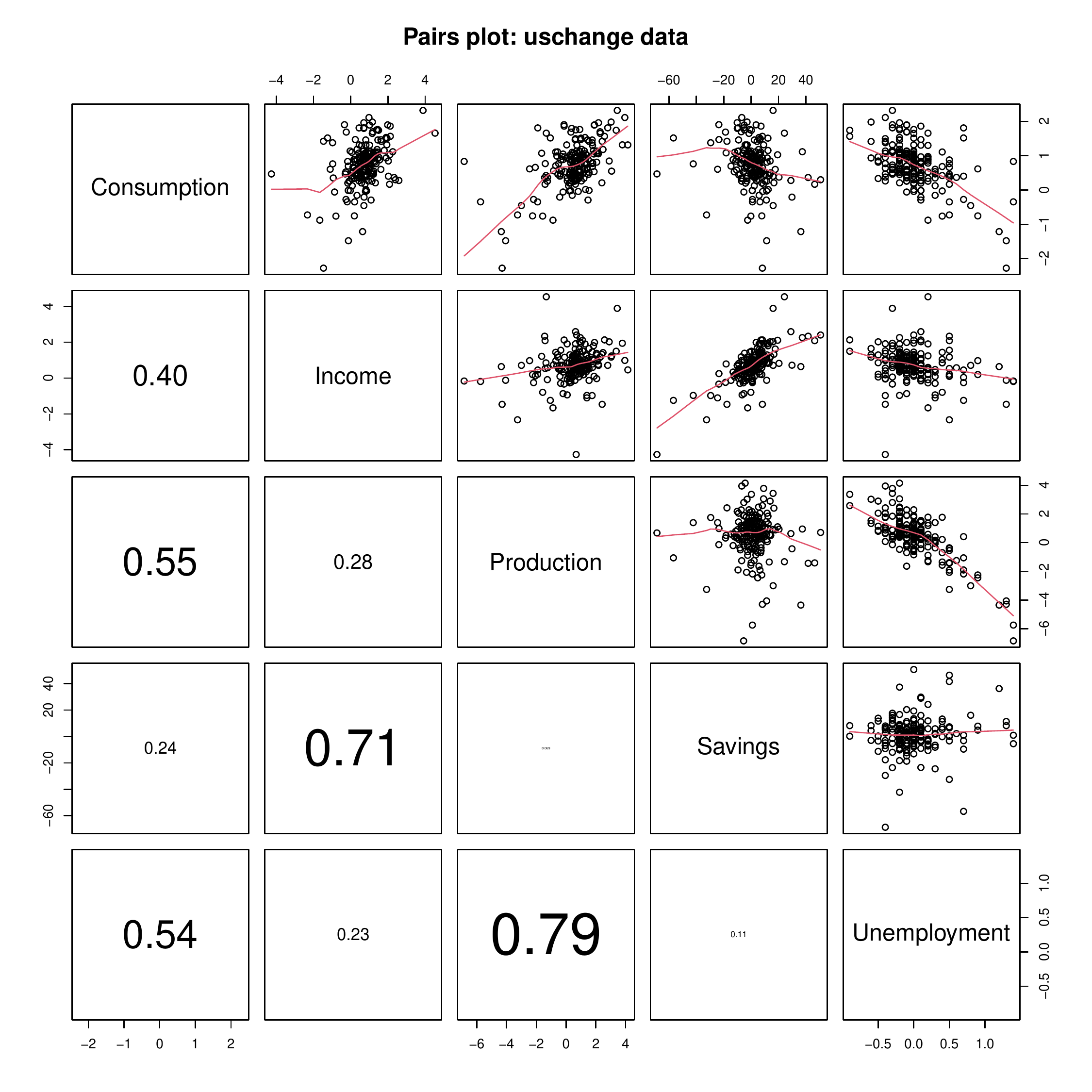}
\caption{\label{fig:USC} Pairs plot of \code{uschange} data with smoothing lines and pair-wise absolute correlations. }
\end{figure}
In total, the data contains 187 observations of 5 variables. The dependency structure is given by the quarters on which the measurements where recorded. The goal is to regress the personal consumption change on all other 4 covariates with an additional intercept. Therefore, the full varying coefficient model in shorthand notation is given by:
\begin{align}
  y_t = \beta_1(t) + \beta_2(t) \texttt{Income}_t + \beta_3(t) \texttt{Production}_t + \beta_4(t) \texttt{Savings}_t + \beta_5(t) \texttt{Unemployment}_t + \varepsilon_t, \label{eq:VCmodel}
\end{align}
with response \code{Consumption}$_t$. In this application the term \emph{spatially} varying coefficient might be confusing and we simply refer to $\beta_j(t)$ as a varying coefficient. The model \eqref{eq:VCmodel} is very similar to the varying-coefficient models introduced by \citet{Hastie1993}. However, our model and methodology differs in the definition and estimation of $\beta_j(\cdot)$. 

\subsubsection{Maximum Likelihood Estimation}

\begin{figure}
\centering
\includegraphics{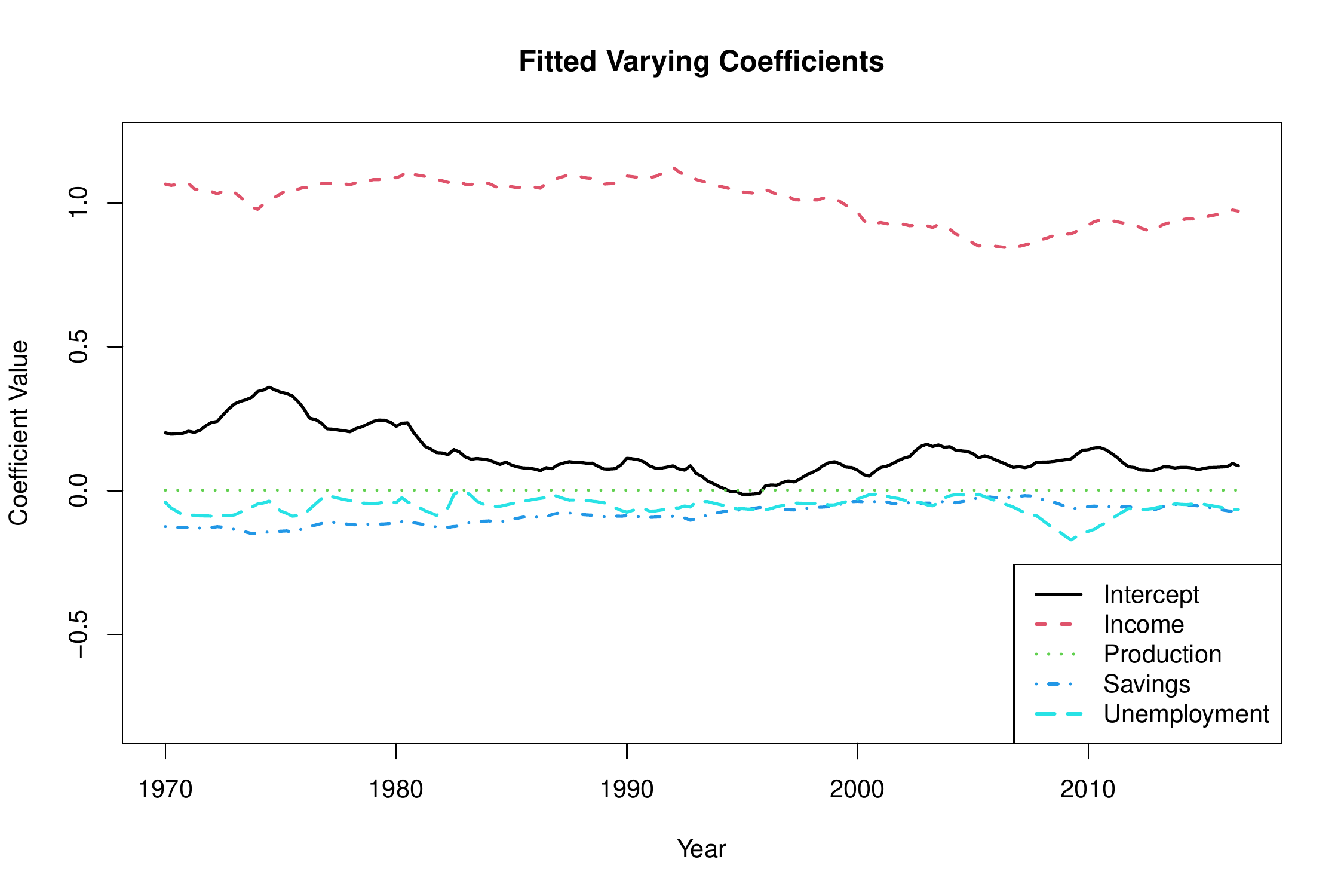}
\caption{\label{fig:USCfittedSVC} Temporal dependency of ML estimated varying coefficients. The estimated mean effects have been added.}
\end{figure}

To determine what kind of model is suitable, i.e., which coefficients are temporally varying and which are not, we will apply the variable selection described above. To this end, we start with a classical MLE of the full varying coefficient model before applying the PMLE.
\begin{Schunk}
\begin{Sinput}
R> # training data
R> X_train <- as.matrix(cbind(Intercept = 1, train_dat[, -1]))
R> y_train <- train_dat$Consumption
R> # prepare cluster: see Examples in ?SVC_mle_control
R> require(parallel)
R> cl <- makeCluster(detectCores()-1, setup_strategy = "sequential")
R> invisible(clusterEvalQ(
+    cl = cl,
+    {
+      library(spam)
+      library(varycoef)
+    }
+  ))
R> # control parameters of MLE including computing cluster
R> control <- SVC_mle_control(
+    profileLik = TRUE, 
+    parallel = list(cl = cl, forward = TRUE, loginfo = TRUE)
+  )
R> SVC_model_mle <- SVC_mle(
+    y = y_train, X = X_train, locs = train_t, control = control
+  )
R> # stop cluster 
R> stopCluster(cl); rm(cl)
R> summary(SVC_model_mle)
\end{Sinput}
\begin{Soutput}
Call:
SVC_mle.default(y = y_train, X = X_train, locs = train_t, control = control)

Fitting a GP-based SVC model with 5 fixed effect(s) and 5 SVC(s)
using 187 observations at 187 different locations / coordinates.

Residuals:
     Min.    1st Qu.     Median    3rd Qu.       Max.  
-0.067810  -0.017306   0.002322   0.014628   0.102556  

Residual standard error: 0.02525
Multiple R-squared: 0.9985, BIC: -250.1


Coefficients of fixed effect(s):
              Estimate Std. Error Z value Pr(>|Z|)    
Intercept     0.130568   0.035166   3.713 0.000205 ***
Income        1.008021   0.033385  30.194  < 2e-16 ***
Production    0.001372   0.006638   0.207 0.836294    
Savings      -0.088688   0.024060  -3.686 0.000228 ***
Unemployment -0.055151   0.051574  -1.069 0.284916    
---
Signif. codes:  0 '***' 0.001 '**' 0.01 '*' 0.05 '.' 0.1 ' ' 1


Covariance parameters of the SVC(s):
                    Estimate Std. Error W value Pr(>W)  
Intercept.range     4.448799   2.191420      NA     NA  
Intercept.var       0.006576   0.002882   5.206 0.0225 *
Income.range        4.109383   1.792707      NA     NA  
Income.var          0.005540   0.002877   3.708 0.0542 .
Production.range    5.028521        NaN      NA     NA  
Production.var      0.000000   0.000000     NaN    NaN  
Savings.range      18.008847  41.658636      NA     NA  
Savings.var         0.001292   0.001761   0.538 0.4632  
Unemployment.range  3.485734        NaN      NA     NA  
Unemployment.var    0.013210        NaN     NaN    NaN  
nugget.var          0.001781   0.001078      NA     NA  
---
Signif. codes:  0 '***' 0.001 '**' 0.01 '*' 0.05 '.' 0.1 ' ' 1

The covariance parameters were estimated using 
exponential covariance functions.
No covariance tapering applied.


MLE:
The MLE terminated after 71 function evaluations with convergence code 52
(0 meaning that the optimization was succesful).
The final profile log likelihood value is 148.6.
\end{Soutput}
\end{Schunk}
Maximum likelihood estimation yields a zero-estimate of the variance of the coefficient of the \code{Production}, i.e., $\hat{\sigma}_3^2(\text{MLE}) = 0$. The fitted varying coefficients are given in Figure~\ref{fig:USCfittedSVC}. Further, we notice that all coefficients apart the intercept's one are sign preserving and have a correctly associated effect. Though, it must be noted that the intercept's dip into negative values is much smaller than the intercept's mean effect standard error. Considering the fitted coefficients, we see that the \code{Unemployment} has some sharp peaks. For instance, $\hat{\beta}_5(t)$ has relatively large volatility in the mid 1970's as well as early 1980's and reaches its minimum for $t\approx 2008$. In all cases, the unemployment in the US reached its all-time highs (prior to the global COVID19 pandemic, c.f. Figure~\ref{fig:USCtime}) due to great recessions like the one caused by the global financial crisis (2007-2008). Finally, we can see that the estimated model has a implausible coefficient of determination of 0.999 which hints at overfitting. Therefore, some regularization or variable selection is necessary.

\subsubsection{Penalized Maximum Likelihood Estimation}

\begin{figure}
\centering
\includegraphics{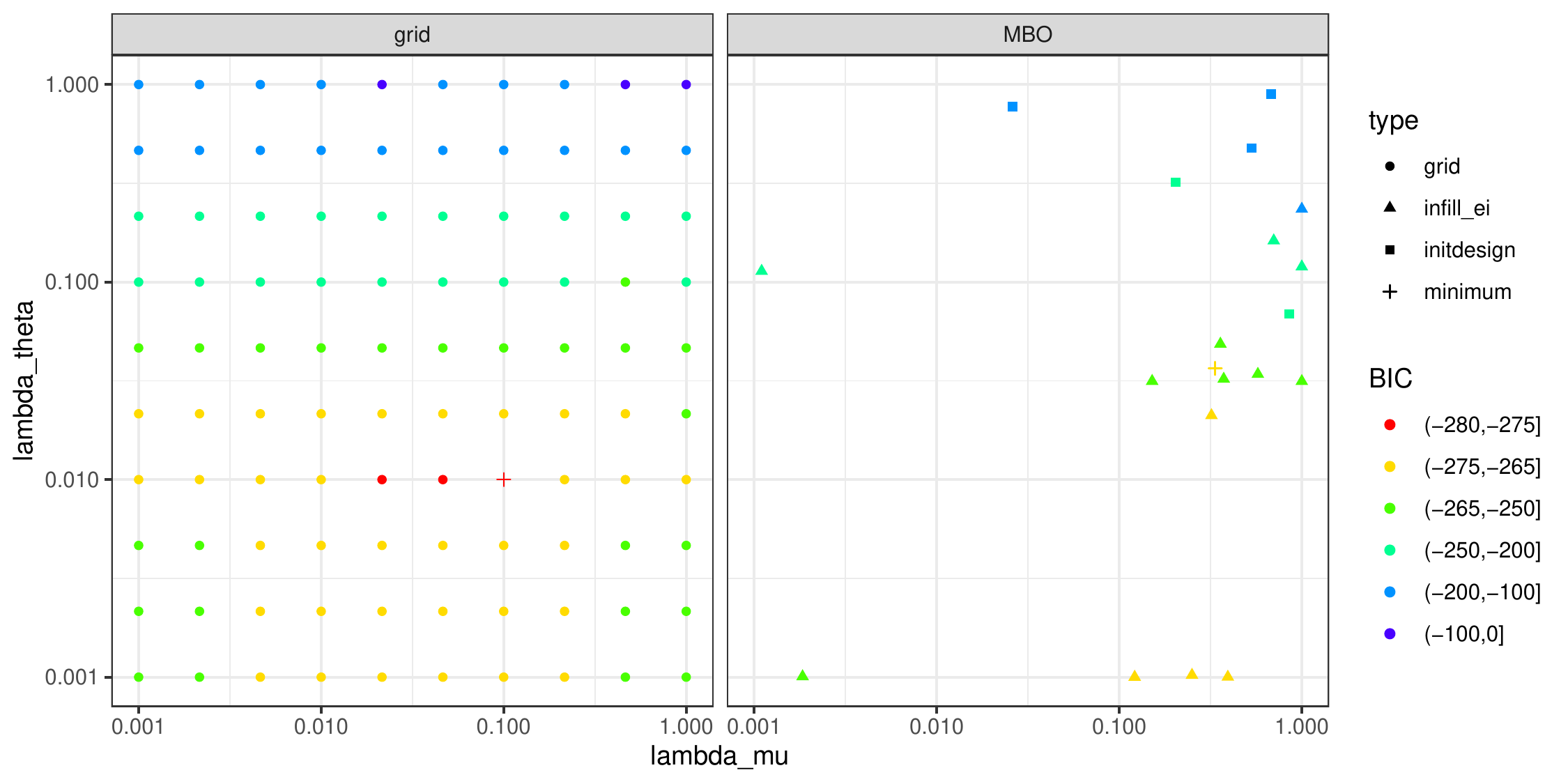}
\caption{\label{fig:USCcompareICO} Comparison of information criterion optimization methods by scatter plots for the computed pairs of $\bigl(\lambda_\bmu, \lambda_\btheta\bigr)$. On the left hand side, we see the $10 \times 10$ grid structure. On the right hand side, we see the MBO. Note that both plots are on a log-log-scale. The MBO's selected shrinkage parameter is part of the initial design. Note that the color-coding for the BIC is deliberately on a non-regular basis to convey more information about the structure of the BIC.}
\end{figure}

In this section we maximize the penalized likelihood and provide respective estimates. In particular, we compare and examine both information criterion optimization methods to obtain the shrinkage parameters. The results for both methods are visualized in Figure~\ref{fig:USCcompareICO}. 
For both methods, the lower and upper bounds of both shrinkage parameters were set to $10^{-3}$ and $1$, respectively. The grid method used a $10 \times 10$ lattice of shrinkage parameters. The MBO method used 5 initial values (squares in Figure~\ref{fig:USCcompareICO}) which span the shrinkage parameter space, before applying a surrogate model with a Gaussian process and updating it after each computation. Here, we used 15 further iterations using the expectation improvement infill criterion (triangles in Figure~\ref{fig:USCcompareICO}). The respective minimums of both methods are given by a plus sign. The selected shrinkage parameters are:
\begin{align*}
  \hat{\blambda}_{grid} = (0.1, 0.01)^\top, \quad \hat{\blambda}_{MBO} = (0.3353, 0.0366)^\top,
\end{align*}
with respective BIC -279.2 for the grid method and -269.8 for the MBO. The ML-estimated GP-based SVC model has a BIC of -250.1. In Figure~\ref{fig:USCcompareICO} we can also see that the BIC surface is relatively flat in the neighborhood of the selected shrinkage parameters due to the small number of observations. In this showcase it is possible that there exist several local minima. The advantage of the MBO is the run time. While the grid method takes 15.7~minutes for the whole variable selection, the MBO requires 4.7~minutes, only 29.9\% of the grid method time.

We present the ML- and PML-estimated parameters of model \eqref{eq:VCmodel} in Table~\ref{tab:uschange-Est}. In terms of variable selection, the initial MLE already has one zero-estimate of the variance for the \code{Production} coefficient. Both methods for optimizing the information criterion yield very similar estimates despite having slightly different selected shrinkage parameters. In both cases, PMLE further increased model sparsity by excluding \code{Production} entirely from the model as well as excluding the random effect for \code{Unemployment}. There are two possible reasons for such behavior. The model over-fitted the data or the Gaussian process was misspecified for instance in the smoothness of the covariance function. In either case the exclusion of the corresponding Gaussian process is the right step as it raises a flag. Further, one can observe that the lack of a time depending \code{Unemployment} coefficient is absorbed by an increase of the error variance.
%
\begin{table}[!t]
\centering
\begin{tabular}{lrrrrrrrrr}
  \hline \textbf{Variable} & \multicolumn{3}{c}{\textbf{Mean} $\hat{\mu}_j$}	&\multicolumn{3}{c}{\textbf{Range} $\hat{\rho}_k$} & \multicolumn{3}{c}{\textbf{Variance} $\hat{\sigma}^2_k$} \\ \cmidrule(lr){2-4} \cmidrule(lr){5-7} \cmidrule(lr){8-10}  & MLE & grid & MBO & MLE & grid & MBO & MLE & grid & MBO  \\ 
  \hline
\texttt{Interc.} & 0.131 & 0.126 & 0.124 & 4.45 & 4.73 & 4.76 & 0.0066 & 0.0062 & 0.0037 \\ 
  \texttt{Income} & 1.008 & 1.013 & 1.014 & 4.11 & 5.05 & 5.15 & 0.0055 & 0.0031 & 0.0017 \\ 
  \texttt{Produc.} & 0.001 & 0.000 & 0.000 & 5.03 & 5.03 & 5.03 & 0.0000 & 0.0000 & 0.0000 \\ 
  \texttt{Savings} & $-$0.089 & $-$0.085 & $-$0.080 & 18.01 & 17.55 & 17.48 & 0.0013 & 0.0013 & 0.0010 \\ 
  \texttt{Unempl.} & $-$0.055 & $-$0.072 & $-$0.063 & 3.49 & 3.24 & 3.24 & 0.0132 & 0.0000 & 0.0000 \\ 
  Error &  &  &  &  &  &  & 0.0018 & 0.0025 & 0.0037 \\ 
   \hline
\end{tabular}
\caption{Parameter estimates for varying coefficient model \eqref{eq:VCmodel} of \code{uschange} data including an intercept (\code{Interc.}). Individual parameters are obtained by MLE and PMLE. For latter, we give parameter estimates for both information criterion optimization methods grid and MBO.} 
\label{tab:uschange-Est}
\end{table}%


\section{Summary} \label{sec:summary}

The \pkg{varycoef} package offers user-friendly tools to model and regress dependent data using (spatially) varying coefficient models. Though the package has been developed with an application to spatial data ($d = 2$) in mind, it now supports the modeling of other types of dependent data, as long as a suitable distance measure between observations exists. The models are flexible with respect to the definition of fixed or random effects and, in particular, with respect to the covariance matrices. Several statistical and computational techniques like parallel computing, covariance tapering including sparse matrix algorithms, and optimization over the profile likelihood have been implemented to foster the applicability to large data sets. Besides model estimation and prediction, our \proglang{R} package offers a variable selection method. Model-based optimization, a powerful computational statistics algorithm, is used in \pkg{varycoef} to make the shrinkage parameter selection computationally efficient.

In this article, we gave a variety of examples that showcase the user-friendly application of \pkg{varycoef} on synthetic and real data. The gain of new insights from the estimated models is substantial and immediate. Here, our package offers a wide variety of \proglang{R} methods such that the usage of Gaussian process-based spatially varying coefficient models is similar to, say, linear models with \fct{lm} or general additive models with \fct{gam} from the package \pkg{mgcv}.


\section*{Computational details}

Most of the presented examples in this paper were generated on a MacBook Pro laptop (macOS~11.4) equipped with an Intel(R) Core(TM) i5-8259U CPU (8 logical processing cores @ 2.3~GHz) and 8 GB of RAM. Results for the Lucas County data set were generated on a Ubuntu server (Ubuntu version~16.04.7) equipped with 8 Intel Xeons E7-2850 (a total of 80 logical processing cores @ 2.0~GHz) and 2 TB of RAM. The intermediate results can be found in the GIT repository \url{https://git.math.uzh.ch/jdambo/jss-paper-open-access}.

The results in this paper were obtained using
\proglang{R}~4.1.0 with the
\pkg{varycoef}~0.3.1 package. \proglang{R} itself and all packages mentioned or used (except for \pkg{INLA}) are available from the Comprehensive \proglang{R} Archive Network (CRAN) at \url{https://CRAN.R-project.org/}. For \pkg{INLA}, please consult \url{https://www.r-inla.org/}.

\section*{Acknowledgments}

JD and FS gratefully acknowledge the support of the Swiss Agency for Innovation \emph{innosuisse} (project number 28408.1~PFES-ES). JD gratefully acknowledges internal publication funding by the Institute for Financial Services Zug (IFZ) of the Lucerne University of Applied Sciences and Arts. RF gratefully acknowledges the support of the Swiss National Science Foundation SNSF-175529.


\bibliography{mybib}


\newpage

\begin{appendix}

\section{Lucas County Data set} 

\subsection{Model Estimation} \label{subsec:SVCModel}

The initial parameters for the optimization are provided by the output of a previous optimization. 
\begin{Schunk}
\begin{Sinput}
R> library(varycoef)
R> ## -- Prepare Data -----
R> # subset of data
R> dat <- house[, c(
+    "price", "yrbuilt", "TLA", "lotsize", "garagesqft", 
+    "stories", "wall", "garage", 
+    "long", "lat"
+  )]
R> # drop two levels of stories due to low frequency
R> dat <- droplevels(
+    dat[!(dat$stories == "two+half" |
+           dat$stories == "three"), ]
+  )
R> # apply log transformations and standardize
R> log.vars <- c("TLA", "lotsize", "garagesqft")
R> for (lv in log.vars) {
+    dat[[paste0("l.", lv)]] <- log(dat[[lv]]+1)
+  }
R> std.vars <- c("yrbuilt", "l.TLA", "l.lotsize", "l.garagesqft")
R> for (sv in std.vars) {
+    dat[[paste0("Z.", sv)]] <- scale(dat[[sv]])
+  }
R> ## -- Build Model, Locations and Response Matrices ------ 
R> # locations (now in kilometers)
R> locs <- dat[, c("long", "lat")]/1000
R> # model matrix 
R> FE_formula <- log(price) ~ 1 + Z.yrbuilt + I(Z.yrbuilt^2) + 
+    Z.l.TLA + Z.l.lotsize + Z.l.garagesqft + 
+    stories + wall + garage
R> # - fixed effects
R> X <- as.matrix(model.matrix(
+    FE_formula, data = dat
+  ))
R> # - random effects (SVC) 
R> W <- as.matrix(X[, 1:6])
R> # response
R> y <- log(dat$price)
R> ## -- initial values and boundaries -----
R> load("~/data-analysis/lucas-county/last_val.RData")
R> init <- as.numeric(last_val)
R> lower <- c(rep(c(0.01, 0), ncol(W)), 1e-6)
R> upper <- c(rep(c(300, 2), ncol(W)), 2)
R> ## -- Prepare MLE ------
R> # start cluster
R> library(parallel)
R> cl <- makeCluster(parallel::detectCores() - 1)
R> clusterEvalQ(cl, {
+    library(spam)
+    library(varycoef)
+  })
R> # control parameters
R> control <- SVC_mle_control(
+    tapering = 1,
+    profileLik = TRUE,
+    init = init, lower = lower, upper = upper,
+    parallel = list(cl = cl, forward = FALSE, loginfo = TRUE)
+  )
R> ## -- Run MLE ------
R> # takes a couple of hours
R> fit <- SVC_mle(
+    y = y, X = X, W = W, 
+    locs = locs, control = control,
+    optim.control = list( 
+      trace = 6, 
+      parscale = abs(ifelse(init == 0, 1, init)))
+  )
R> # stop Cluster
R> stopCluster(cl); rm(cl)
\end{Sinput}
\end{Schunk}

\subsection{Remaining Estimated SVCs} \label{subsec:remainSVC}

\begin{figure}
\centering
\includegraphics{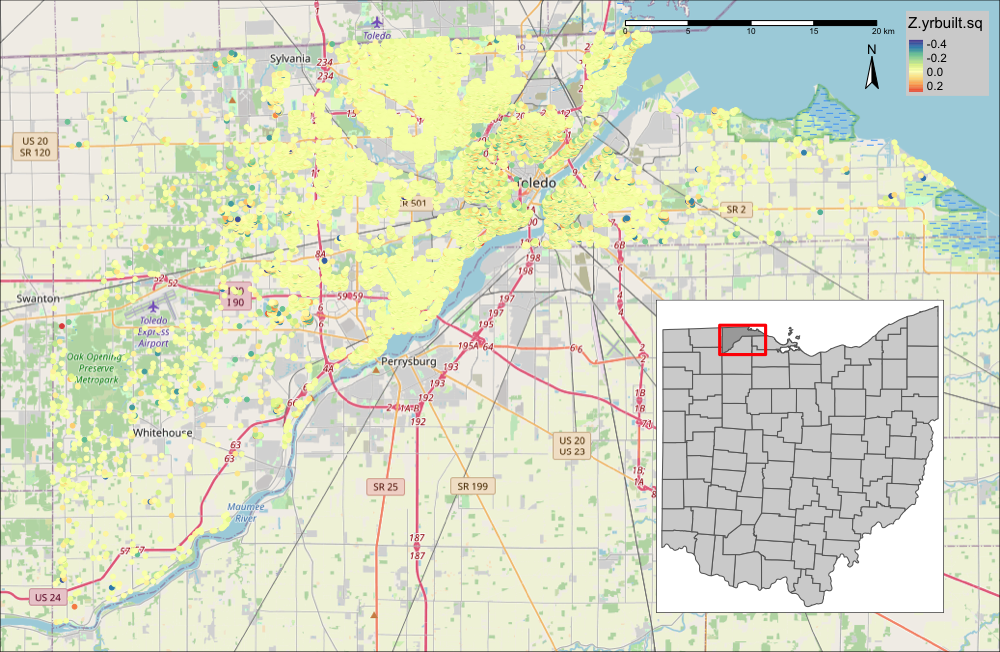}
 \vspace{1cm}
 \includegraphics{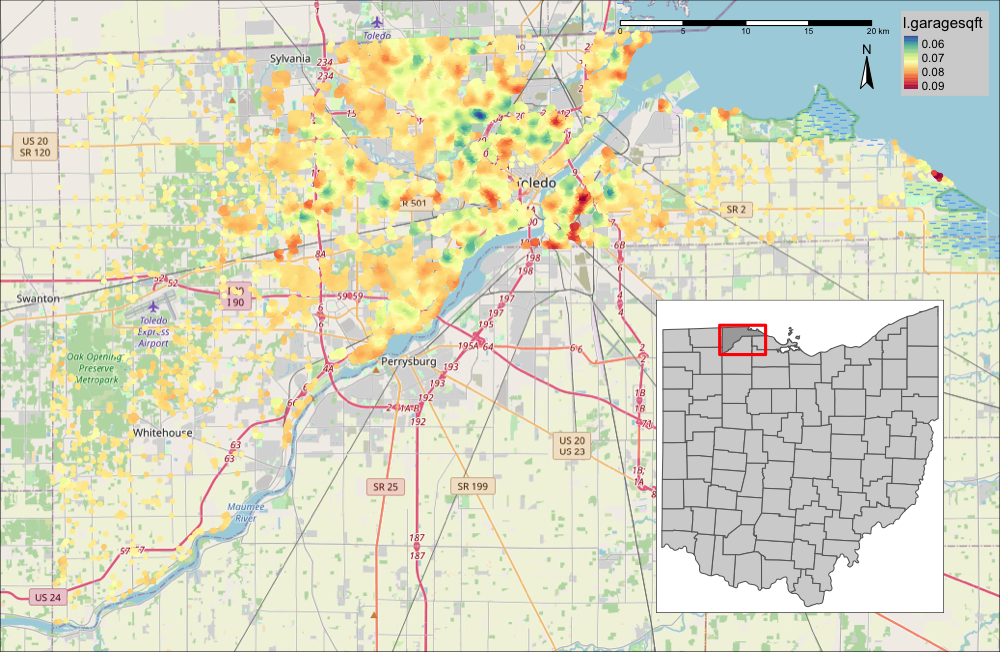}
\caption{\label{fig:SVC-LC2} Third and sixth estimated SVC.}
\end{figure}

In Figure~\ref{fig:SVC-LC2}, we present the remaining estimated spatially varying coefficients of the GP-based SVC model for the Lucas County data. Note that the coefficients for the total living area and the lot size are flat.

\newpage

\section{US Growth Rate Data set} \label{app:uschange}

The code below runs the PMLE on the \code{uschange} data. The results were mentioned in Section~\ref{sec:uschange}. 
\begin{Schunk}
\begin{Sinput}
R> ## -- US Change -------
R> library(fpp2)
R> data("uschange")
R> # divide times series data into...
R> str(uschange)
R> # ... variables and ...
R> train_dat <- as.data.frame(uschange)
R> # ... time points of observations
R> train_t <- as.numeric(time(uschange))
R> # visual inspection
R> autoplot(uschange, facet = TRUE)
R> pairs(train_dat)
R> # classical linear model
R> linmod <- lm(Consumption~., dat = train_dat)
R> summary(linmod)
R> ## -- MLE ------
R> # prepare for SVC model
R> X_train <- model.matrix(linmod)
R> y_train <- train_dat$Consumption
R> p <- ncol(X_train)
R> require(parallel)
R> cl <- makeCluster(detectCores()-1)
R> clusterEvalQ(
+    cl = cl,
+    {
+      library(spam)
+      library(varycoef)
+    })
R> # use this list for parallel argument in SVC_mle_control
R> parallel.control <- list(cl = cl, forward = TRUE, loginfo = TRUE)
R> # SVC Modeling
R> library(varycoef)
R> control <- SVC_mle_control(profileLik = TRUE, parallel = parallel.control)
R> SVC_model_mle <- SVC_mle(
+    y = y_train, X = X_train, locs = train_t, 
+    control = control, optim.control = list(trace = 6)
+  )
R> summary(SVC_model_mle)
R> ## -- PMLE -----
R> control$extract_fun <- TRUE
R> obj_fun <- SVC_mle(
+    y = y_train, 
+    X = X_train, 
+    W = X_train, 
+    locs = train_t,
+    control = control
+  )
R> ## grid
R> # set controls for SVC selection
R> sel_control1 <- SVC_selection_control(
+    method = "grid",
+    IC.type = "BIC",
+    r.lambda = c(1e-3, 1),
+    n.lambda = 10L, 
+    CD.conv = list(N = 20, delta = 1e-8, logLik = FALSE),
+    parallel = control$parallel,
+    optim.args = list(
+      lower = SVC_model_mle$MLE$comp.args$liu$lower,
+      upper = SVC_model_mle$MLE$comp.args$liu$upper
+    ),
+    adaptive = TRUE
+  )
R> # run PMLE
R> time_start1 <- Sys.time()
R> PMLE_CD1 <- SVC_selection(
+    obj_fun, mle.par = cov_par(SVC_model_mle),
+    control = sel_control1,
+    approx = FALSE
+  )
R> time_end1 <- Sys.time()
R> ## MBO
R> sel_control2 <- SVC_selection_control(
+    method = "MBO",
+    IC.type = "BIC",
+    r.lambda = c(1e-3, 1),
+    n.init = 5L,
+    n.iter = 15L,
+    CD.conv = list(N = 20, delta = 1e-8, logLik = FALSE),
+    parallel = control$parallel,
+    optim.args = list(
+      lower = SVC_model_mle$MLE$comp.args$liu$lower,
+      upper = SVC_model_mle$MLE$comp.args$liu$upper
+    ),
+    adaptive = TRUE
+  )
R> # run PMLE
R> time_start2 <- Sys.time()
R> set.seed(1)
R> PMLE_CD2 <- SVC_selection(
+    obj_fun, mle.par = cov_par(SVC_model_mle),
+    control = sel_control2,
+    approx = FALSE
+  ) 
R> time_end2 <- Sys.time()
R> # timings
R> df_timings <- data.frame(
+    method = c("grid", "MBO"), 
+    t_start = c(time_start1, time_start2), 
+    t_end = c(time_end1, time_end2)
+  )
R> # save PMLEs
R> save(PMLE_CD1, PMLE_CD2, df_timings,
+       file = "data-analysis/uschange/sel-outcomes.RData")
\end{Sinput}
\end{Schunk}

\end{appendix}


\end{document}